\documentclass[aps,prb,showpacs,twocolumn,floats]{revtex4}
\usepackage{amssymb}
\usepackage{amsbsy}
\usepackage{amsmath}
\usepackage{graphicx}
\usepackage{amsmath}
\usepackage{times}
\usepackage{color}
\usepackage{subfigure}
\usepackage{setspace}
\usepackage{bm}% bold math
\newcommand\bea{\begin{eqnarray}}
\newcommand\eea{\end{eqnarray}}
\newcommand\beq{\begin{equation}}
\newcommand\eeq{\end{equation}}
\newcommand\bib{\bibitem}

\begin{document}

\title{Josephson junctions of Weyl and multi-Weyl semimetals}

\author{Kirill Kulikov$^1$, Debabrata Sinha$^2$, Yu. M. Shukrinov$^{1,3}$, and K. Sengupta$^4$}
\affiliation{$^1$ BLTP, Joint Institute for Nuclear Research, Dubna,
Moscow Region, 141980, Russia \\
$^2$ Center for Theoretical Studies, Indian Institute of Technology,
Kharagpur-721302, India \\
$^3$ Department of Physics, Dubna State University, Dubna, 141980, Russia\\
$^4$ School of Physical Sciences, Indian Association for the
Cultivation of Science, Jadavpur, Kolkata-700032, India.}

\date{\today}

\begin{abstract}

We study a Josephson junction involving a Weyl and a multi-Weyl
semimetal separated by a barrier region of width $d$ created by
putting a gate voltage $U_0$ over the Weyl semimetal. The
topological winding number of such a junction changes across the
barrier. We show that $I_c R_N$ for such junctions, where $I_c$ is
the critical current and $R_N$ the normal state resistance, in the
thin barrier limit, has a universal value independent of the barrier
potential. We provide an analytical expression of the Andreev bound
states and use it to demonstrate that the universal value of $I_c
R_N$ is a consequence of change in topological winding number across
the junction. We also study AC Josephson effect in such a junction
in the presence of an external microwave radiation, chart out its
current-voltage characteristics, and show that the change in the
winding number across the junction shapes the properties of its
Shapiro steps. We discuss the effect of increasing barrier thickness
$d$ on the above-mentioned properties and chart out experiments
which may test our theory.

\end{abstract}


\maketitle

\section{Introduction}
\label{int}

The role of topology in shaping the low energy properties of
condensed matter systems has received widespread attention in recent
years \cite{rev1,rev2}. Specific examples of such materials include
graphene \cite{rev3}, topological insulators \cite{rev4}, and Dirac
and Weyl semimetals \cite{rev2}. These are classes of materials
whose low-energy quasiparticles obey Dirac or Weyl equations and
host Dirac/Weyl nodes at specific points in their Brillouin zone.
These nodes act as sources or sinks of topological winding numbers.
Several unconventional low-energy properties of these materials
arise from the presence of such nodes. For example, in graphene, the
magnetoresistivity of electrons displays an additional contribution
due to the non-zero Berry phase gathered by electrons whose
semiclassical trajectories encompasses a Dirac node \cite{exp1}. For
topological insulators, the surface quasiparticles obey
spin-momentum locking as seen in spin- and angle-resolved
photoemission experiments \cite{exp2}. In three-dimensional (3D)
Weyl semimetals, the presence of such nodes leads to a host of
unconventional phenomena such as Fermi arcs on their surface
\cite{arcref}, negative magnetoresistance \cite{magneref}, chiral
anomaly \cite{anmref}, and interaction induced phase transitions
\cite{br1}

A typical Weyl semimetal hosts several Weyl nodes in its Brillouin
zone. The effective low-energy of quasiparticles near any of such
Weyl nodes is given by $H= \pm \hbar v_F \sum_{\vec k} {\vec \sigma
} \cdot {\vec k}$, where $\vec k= (k_x, k_y, k_z)$ denotes momentum
measured from the Weyl node, $v_F$ is the Fermi velocity, the +(-)
sign corresponds to nodes with positive(negative) chirality, and
$\vec \sigma = (\sigma_x, \sigma_y, \sigma_z)$ are Pauli matrices in
spin-space. Such Weyl nodes, occurring in pairs, are usually
protected by either time-reversal or inversion symmetry. More
recently, it was suggested that certain materials such as Hg${\rm
Te}_2$ may host Weyl nodes with anisotropic dispersion along two
transverse directions (denoted as $\vec k_t = (k_x, k_y)$ in this
work): $E \sim \hbar v_F \sqrt{k_z^2 + \epsilon_0^2 |\vec
k_t|^{2n}}$, where $\epsilon_0$ is a constant whose value depends on
the details of band dispersion in the material and $n \ne 1$ is an
integer indicating the anisotropy. It was shown that such a
dispersion arises out of symmetry protected (such as discrete three-
or six-fold rotational symmetry) merger of $n \le 3$ Weyl nodes
\cite{multi1}. These materials are dubbed as multi-Weyl semimetals
and are known to have several unconventional properties which are
distinct from both conventional metals and Weyl semimetals
\cite{multi2}. In particular, these multi-Weyl nodes have a
topological winding number $3 \ge n > 1$ ($n \in Z$) which is also
the number of Weyl nodes merged. Interestingly, an application of
external strain, which lifts the rotational symmetry protecting the
merger, leads to emergence of $n$ Weyl nodes of same chirality and
unit winding number from a multi-Weyl node.

The study of ballistic transport in junctions of materials whose
low-energy quasiparticles satisfy Dirac or Weyl equations provides a
wealth of information regarding their topological properties. In
addition, it also provides access to several unconventional
transport behavior displayed by these materials. For example, in 2D
materials with Dirac quasiparticles (such as graphene and surfaces
of topological insulators), the tunneling conductance $G$ of
normal-barrier-superconductor (NBS) junctions oscillates as a
function of barrier potential \cite{ks1,been1}; such a behavior is
qualitatively distinct from the monotonic decay of $G$ with
increasing barrier potential in conventional junctions. A similar
oscillatory dependence has been reported for junctions of Weyl
semimetals \cite{weyltrans1,ds1}. In contrast, it was found that
normal-barrier-normal (NBN) and normal-barrier-superconducting (NBS)
junctions of Weyl and multi-Weyl semimetals display barrier
independence of $G$ in the thin barrier limit \cite{ds1}.

For 2D Dirac materials an analogous oscillatory dependence of the
Josephson current $I_J$ on the barrier potential has also been
predicted for superconductor-barrier-superconductor (SBS) junctions
\cite{ks2,been2,yu1}. It is well known that for conventional SBS
junctions, the product $I_c R_N$, where $I_c$ is the critical
current and $R_N$ is the normal state resistance of the junction, is
a monotonically decreasing function barrier strength. The value of
$I_c R_n/[\pi \Delta_0 /(2e)]$ (where $\Delta_0$ is the amplitude of
the pair-potential $\Delta$ and $e$ is the electron charge)
decreases from $2$ to $1$ in conventional Josephson junctions with
increasing barrier strength for thin barriers. Its maximal value
occurs in the Kulik-Omelyanchuk (KO) limit when the junction is
transparent \cite{ko1} while the minimum value is reached for
junctions with very large value of the barrier potential in the
so-called Ambegaokar-Baratoff (AB) limit \cite{ab1}. In contrast,
SBS junctions whose quasiparticles obey Dirac-Bogoliubov-de Gennes
(DBdG) equations never reach the AB limit \cite{ks2}; the value of
$I_cR_n/[\pi \Delta_0/(2e)]$ in these junctions oscillates as a
function of barrier strength between $2$ and $1.13$ \cite{ks2}. A
similar oscillatory behavior is observed in junctions of Weyl
semimetals \cite{weyltrans1}. However, the properties of AC
Josephson effect in the presence of microwave radiation has not been
investigated for these junctions. Moreover, either DC or AC
Josephson effect between junctions of Weyl and multi-Weyl semimetals
has not been studied so far.

In this work, we study the DC and AC Josephson effect in SBS
junctions between Weyl and multi-Weyl semimetals. The main results
that we obtain are as follows. First, we find analytic expressions
for Andreev bound states in these junctions in the thin barrier
limit. In this limit, the barrier potential $V \to \infty$ and its
thickness $d \to 0$ such that the dimensionless barrier strength
$\chi= U_0 d/(\hbar v_F)$, where $v_F $ is the Fermi velocity of the
DBdG quasiparticles and $\hbar$ is Planck's constant, is finite.
Using the analytic expression Andreev bound states, we also obtain
expressions for both DC and AC Josephson current in such junctions.
Second, we show that $I_c R_N$ in these junctions is a barrier
independent constant in the thin barrier limit: $I_c R_N = c [\pi
\Delta_0/(2e)]$ for any $\chi$. Here $c$ is a constant which depends
weakly on $n_1$ and $n_2$; numerically we find $c=1.56$ for $n_1=1$
and $n_2=2$ and $c=1.62$ for $n_1=1$ and $n_2=3$. We demonstrate
that this behavior, which is contrast to behavior of $I_c R_N$ in
all Josephson junctions studied earlier, is a consequence of change
in the topological winding number across the junction and estimate
the deviation of $I_c R_N$ from its thin barrier limit as a function
of $U_0$ for thicker barriers. Third, we study the AC Josephson
effect in such junctions in the presence of a microwave radiation
both with voltage and current biases. We obtain the current-voltage
(I-V) characteristics of such junctions and demonstrate the presence
of Shapiro steps in the I-V curves. We show that the width of these
Shapiro steps for such junctions is independent of the barrier
strength in the thin barrier limit. Finally, we suggest possible
experiments which may test our theory.

The plan of the rest of this work is as follows. In Sec.\ \ref{and1}
we chart out the derivation of Andreev bound states and use it show
the barrier independence of $I_c R_N$. This is followed by Sec.\
\ref{acj}, where we discuss AC Josephson effect in such junctions.
Finally, in Sec.\ \ref{conc}, we summarize our main results, discuss
experiments which may test our theory, and conclude.

\section{Andreev Bound states}
\label{and1}

\begin{figure}
\rotatebox{0}{\includegraphics*[width=\linewidth]{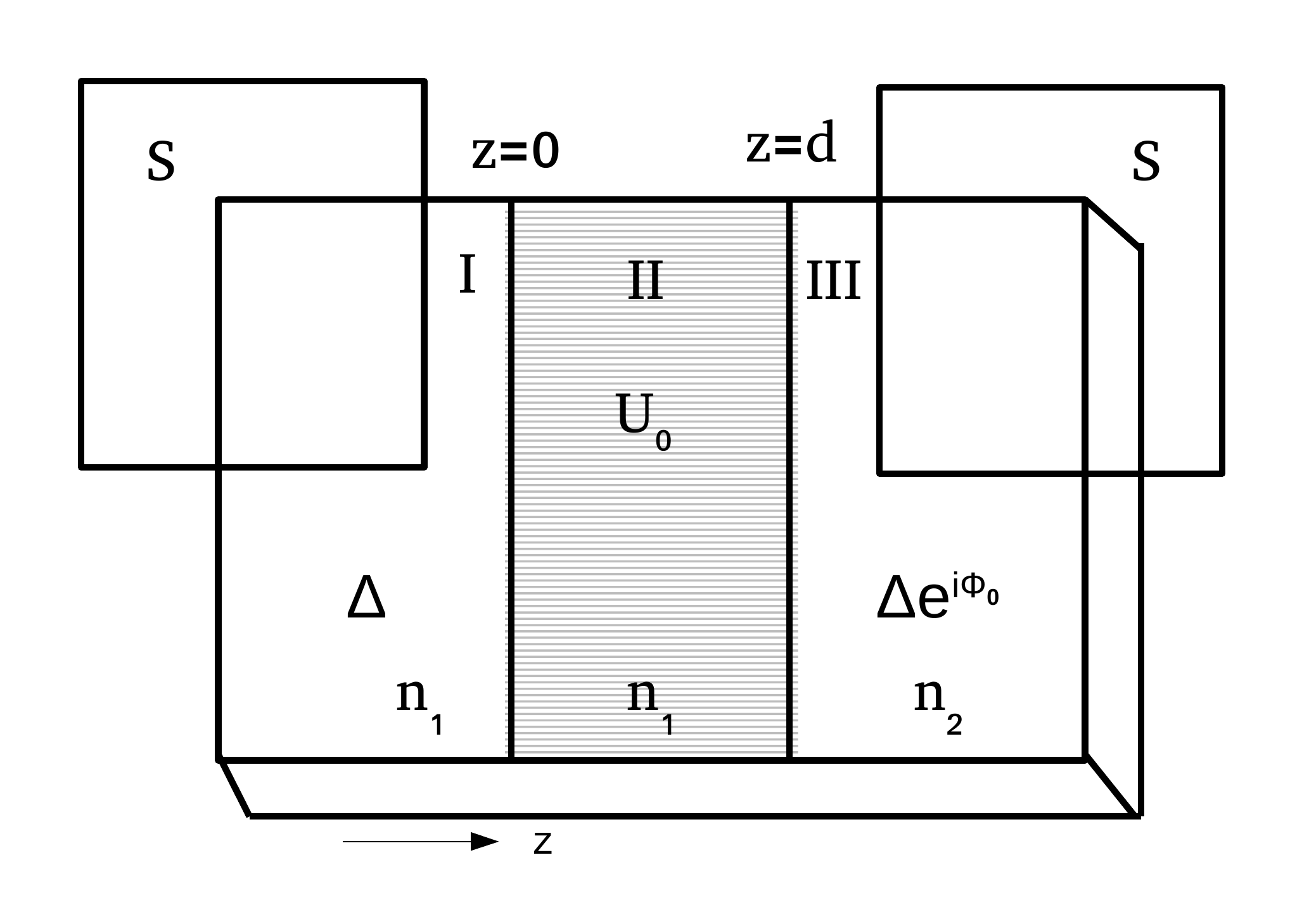}}
\caption{A schematic representation of the Josephson junction
geometry. The barrier region involves a gate potential $U_0$ applied
in region II and extends from $z=0$ to $z=d$. The proximate
superconductors atop regions I and III induce $s$-wave
superconductivity in these regions; the corresponding pair
potentials $\Delta=\Delta_0$ (region I) and $\Delta=\Delta_0 e^{i
\phi_0}$ (region III) in these two regions have a relative phase
$\phi_0$. The topological winding number corresponding to the
Weyl/muti-Weyl cones is $n_1$ in regions I and II, and $n_2$ in
region III. See text for details. } \label{fig1}
\end{figure}

In this section, we develop analytic expressions for Andreev bound
states in SBS junctions involving a Weyl and a multi-Weyl semimetal
and use them to obtain their Josephson current. In what follows, we
consider a ballistic junction schematically shown in Fig.\
\ref{fig1}. The barrier region has width $d$ and is created by
putting a barrier potential over a normal Weyl semimetal. In
contrast, regions I and III have s-wave superconducting
pair-potentials of amplitude $\Delta_0$: $\Delta=\Delta_0 \exp[i
\phi]$, where $\phi$ is the global superconducting phase. In what
follows we shall choose this phase to be zero in region I and
$\phi_0$ in region III without loss of generality. We assume that
superconductivity has been induced in these regions by proximate
$s$- wave superconductors. The microscopic analysis leading to
specific conditions for $s-$ wave pair potentials for induced
superconductivity in these materials has been carried out in Refs.\
\onlinecite{weyltrans1, multiweylsup1}; here we shall assume that
these conditions hold for our system. In what follows, we shall
consider the situation, where there are two Weyl nodes of opposite
chirality in regions I and III; the winding numbers corresponding to
these nodes will be denotes as $n_1(n_2)$ in regions I (III). The
DBdG quasiparticles of the superconductors arises from superposition
of electrons in one of these nodes with holes in the other. We shall
obtain the equations leading to expression of Andreev bound state
for arbitrary $d$ and $U_0$ in these junctions; the analytic
expression of these bound states will be presented in the
thin-barrier limit as a function of $\chi$.

The low-energy effective Hamiltonian of a multi-Weyl semimetal with
a topological winding number $n$ is given by $ H = \sum_{\vec k}
\Psi_{\vec k}^{\dagger} H_1[n] \Psi_{\vec k}$, where $\Psi_{\vec k}=
(c_{\vec k \uparrow}, c_{\vec k \downarrow})^T$ are two component
fermions field, $c_{\vec k \sigma}$ denotes annihilation operator of
a Weyl quasiparticle with momentum $\vec k$ and spin $\sigma$, and
$H_1[n]$ is given by \cite{multi1}
\begin{eqnarray}
H_1[n] &=&  \left[ \hbar v_F k_z \sigma_z + \epsilon_0 |\vec
k_t|^{n}
\left( \cos(n \phi_{\vec k}) \sigma_x \right. \right. \nonumber\\
&& \left. \left. + i \sin(n \phi_{\vec k}) \sigma_y \right) -\mu_0 I
\right] \label{normham1}
\end{eqnarray}
Here $v_F $ is the Fermi velocity , $k_F=\mu_0/(\hbar v_F)$ is the
Fermi momentum, $\mu_0$ is the chemical potential, $I$ denotes $2
\times 2$ identity matrix, $ \epsilon_0 = \mu_0/k_F^{n}$ is a
material dependent constant (chosen to be unity if $n=1$) whose
numerical value is unimportant for our analysis, and $\phi_{\vec k}
= \arctan[k_y/k_x]$ is the azimuthal angle in the transverse
direction. In the presence of induced $s$-wave superconductivity,
the low-energy Hamiltonian governing the DBdG quasiparticles are
given by
\begin{eqnarray}
H_s[n] &=& \sum_{\vec k} \Psi_{\vec k}^{' \dagger} \left(H_1[n]
\tau_z + \Delta \tau_+ +\Delta^{\ast} \tau_- \right) \Psi'_{\vec k},
\label{supham}
\end{eqnarray}
where $\vec \tau=(\tau_x, \tau_y, \tau_z)$ denotes Pauli matrices
valley space and $\tau_{\pm} = (\tau_x \pm i \tau_y)/2$. Here the
pair potential connects electrons and holes between two Weyl nodes
of opposite chirality (two valleys) and $\Psi'_{\vec k}=(c_{k
\sigma}, c_{-\vec k \bar \sigma}^{\dagger})^T$ denotes the
four-component fermionic fields. In our notation the electrons and
holes of these four components wavefunctions belongs to two
different nodes (valleys).

The Hamiltonian in region I (see Fig.\ \ref{fig1}) is given by
$H_s[n=n_1]$. In what follows, we shall choose $z$ to be the
longitudinal direction. Thus the Weyl equation whose solution yields
left moving electron-like and hole-like DbDG quasiparticles in
region I is given by
\begin{eqnarray}
H_s[n_1; k_z \to -i \partial_z] \psi = E \psi \label{weyl1}
\end{eqnarray}
These wavefunctions are given by \cite{ds1,weyltrans1}
\begin{widetext}
\begin{eqnarray}
\psi_{\rm elq} &=& \left( \sin(\theta_{\vec k 1})e^{i\gamma_1},
\cos(\theta_{\vec k 1}) e^{i \gamma_1},\sin(\theta_{\vec k 1}),
\cos(\theta_{\vec k 1}) \right) e^{i (-k_z^{s(1)} z + \vec k_t \cdot
\vec
r_{t} - n_1 \phi_{\vec k} \sigma_z /2) }/\sqrt{2} \nonumber\\
\psi_{\rm hlq} &=& \left( \cos(\theta_{\vec k 2}), \sin(\theta_{\vec
k 2}, e^{i \gamma_1} \cos(\theta_{\vec k 2}), e^{i \gamma_1}
\sin(\theta_{\vec k 2}) \right) e^{i (k_z^{s(2)} z + \vec k_t \cdot
\vec r_{t} - n_1 \phi_{\vec k} \sigma_z /2) }/\sqrt{2},
\label{wavfn1}
\end{eqnarray}
\end{widetext}
where $k_z^{s (1) [(2)]}= +[-] \sqrt{(\mu_0 +[-] i
\zeta)^2-\epsilon_0^2 |\vec k_t|^{2n_1}}/(\hbar v_F)$, $\zeta =
\sqrt{\Delta_0^2 - E^2(\vec k_t)}$, $\gamma = \arccos(E(\vec
k_t)/\Delta_0)$, and $\tan(2\theta_{\vec k 1[2]})= \epsilon_0 |{\vec
k}_t|^{n}|/(\hbar v_F k_z^{s (1)[(2)]})$. Here we have set the phase
of the superconducting pair-potential in region I to zero without
loss of generality and $\vec r_t = (x,y)$. Note that the dependence
on $\phi_{\vec k}$ of these wavefunctions is equivalent to a
rotation in spin-space by $n_1\phi_{\vec k}$ about the $z-$axis
\cite{ds1}. The wavefunction in region I is thus given by
\begin{eqnarray}
\psi_{I} &=& a_1 \psi_{\rm elq} + b_1 \psi_{\rm hlq}, \label{wavfn2}
\end{eqnarray}
where $a_1$ and $b_1$ are arbitrary complex coefficients.

In region II, the Hamiltonian is given by $H$ with $n=n_1$. The
wavefunction in region II is thus a superposition of left and right
moving electrons and holes. The wavefunction of these quasiparticles
in the four component notation is given by
\begin{eqnarray}
\psi_{e +} &=& (\cos \theta_{\vec k 0}, \sin \theta_{\vec k
0},0,0)e^{ i(k_z z + \vec k_t \cdot \vec r_t - n_1 \sigma_z
\phi_{\vec k}/2)}
\nonumber\\
\psi_{e -} &=& (\sin \theta_{\vec k 0}, \cos \theta_{\vec k
0},0,0)e^{ i(-k_z z + \vec k_t \cdot \vec r_t - n_1 \sigma_z
\phi_{\vec k}/2)}
\nonumber\\
\psi_{h +} &=& (0,0,-\sin \theta'_{\vec k 0}, \cos \theta'_{\vec k
0})e^{ i(k'_z z + \vec k_t \cdot \vec r_t - n_1 \sigma_z \phi_{\vec
k}/2)}
\nonumber\\
\psi_{h -} &=& (0,0,\cos \theta'_{\vec k 0}, -\sin \theta'_{\vec k
0})e^{ i(-k'_z z + \vec k_t \cdot \vec r_t - n_1 \sigma_z \phi_{\vec
k}/2)} ,\label{reg2wav1}
\end{eqnarray}
where $\sin \theta_{\vec k 0} ={\rm Sgn}(E+\mu_0-U_0) |\vec
k_t|^{n_1} \epsilon_0/(E+\mu_0-U_0)$ and $\sin \theta'_{\vec k 0} =
{\rm Sgn}(E-\mu_0+U_0) |\vec k_t|^{n_1} \epsilon_0/(E-\mu_0+U_0)$
are the angles of propagation for electrons and holes respectively
in region II and ${\rm Sgn}$ denotes the signum function. Here $k_z
[k'_z] = {\rm Sgn}(E+[-]\mu_0 -[+] U_0) \sqrt{(E+[-]\mu_0-[+]U_0)^2-
\epsilon_0^2 |{\vec k}_t|^{2 n_1}}/(\hbar v_F)$, and the indices
$+(-)$ denotes right(left) moving electrons and holes. The
wavefunction in region II is given by superposition of these left
and right moving electron and hole wavefunctions and is given by
\begin{eqnarray}
\psi_{II} &=& p \psi_{e+} + q \psi_{e-} + r \psi_{h+} + s \psi_{h -},
\label{reg2wav2}
\end{eqnarray}
where $p$, $q$, $r$, and $s$ are complex coefficients.

In region III, the Hamiltonian is given by $H_s[n=n_2]$ with $\mu_0
\to \mu'_0$. The change in chemical potential between the two
regions is kept to point out the generality of the procedure;
however, it is to be noted that in principle the chemical potentials
in region I and III can always be made equal by applying a gate
voltage between the two regions. The wavefunction in this region is
again a superposition of electron- and hole-like DBdG quasiparticle
wavefunctions. These wavefunctions are given by
\begin{widetext}
\begin{eqnarray}
\psi'_{\rm elq} &=& \left( \cos(\theta_{\vec k 3})e^{i\gamma_1},
\sin(\theta_{\vec k 3}) e^{i \gamma_1},\cos(\theta_{\vec k 3})e^{i
\phi_0}, \cos(\sin_{\vec k 3}) e^{i \phi_0}\right) e^{i (-k_z^{s(3)}
z + \vec k_t \cdot \vec
r_{t} - n_2 \phi_{\vec k} \sigma_z /2) }/\sqrt{2} \nonumber\\
\psi'_{\rm hlq} &=& \left( \sin(\theta_{\vec k 4})e^{i \phi_0},
\cos(\theta_{\vec k 4} e^{i \phi_0}, e^{i \gamma_1}
\sin(\theta_{\vec k 4}), e^{i \gamma_1} \cos(\theta_{\vec k 4})
\right) e^{i (k_z^{s (4)} z + \vec k_t \cdot \vec r_{t} - n_2
\phi_{\vec k} \sigma_z /2) }\sqrt{2}, \label{wavfn3}
\end{eqnarray}
\end{widetext}
where $\phi_0$, as defined earlier, is the relative phase between
the superconductors in region I and III. Here the expressions for
$k_{z}^{s(3)}$ and $k_z^{s(4)}$ may be obtained from those of
$k_z^{s(1)}$ and $k_z^{s(2)}$ respectively by replacing $\mu_0 \to
\mu_0'$ and $n_1 \to n_2$. Similarly $\theta_{\vec k 3(4)}$ can be
read off from the expressions of $\theta_{\vec k 1(2)}$ after making
similar replacements. The wavefunction in region III can then be
written as
\begin{eqnarray}
\psi_{III} &=& a_2 \psi'_{\rm elq} + b_2 \psi'_{\rm hlq},
\label{wavfn4}
\end{eqnarray}
where $a_2$ and $b_2$ are complex coefficients.

\begin{figure}
\rotatebox{0}{\includegraphics*[width=0.48\linewidth]{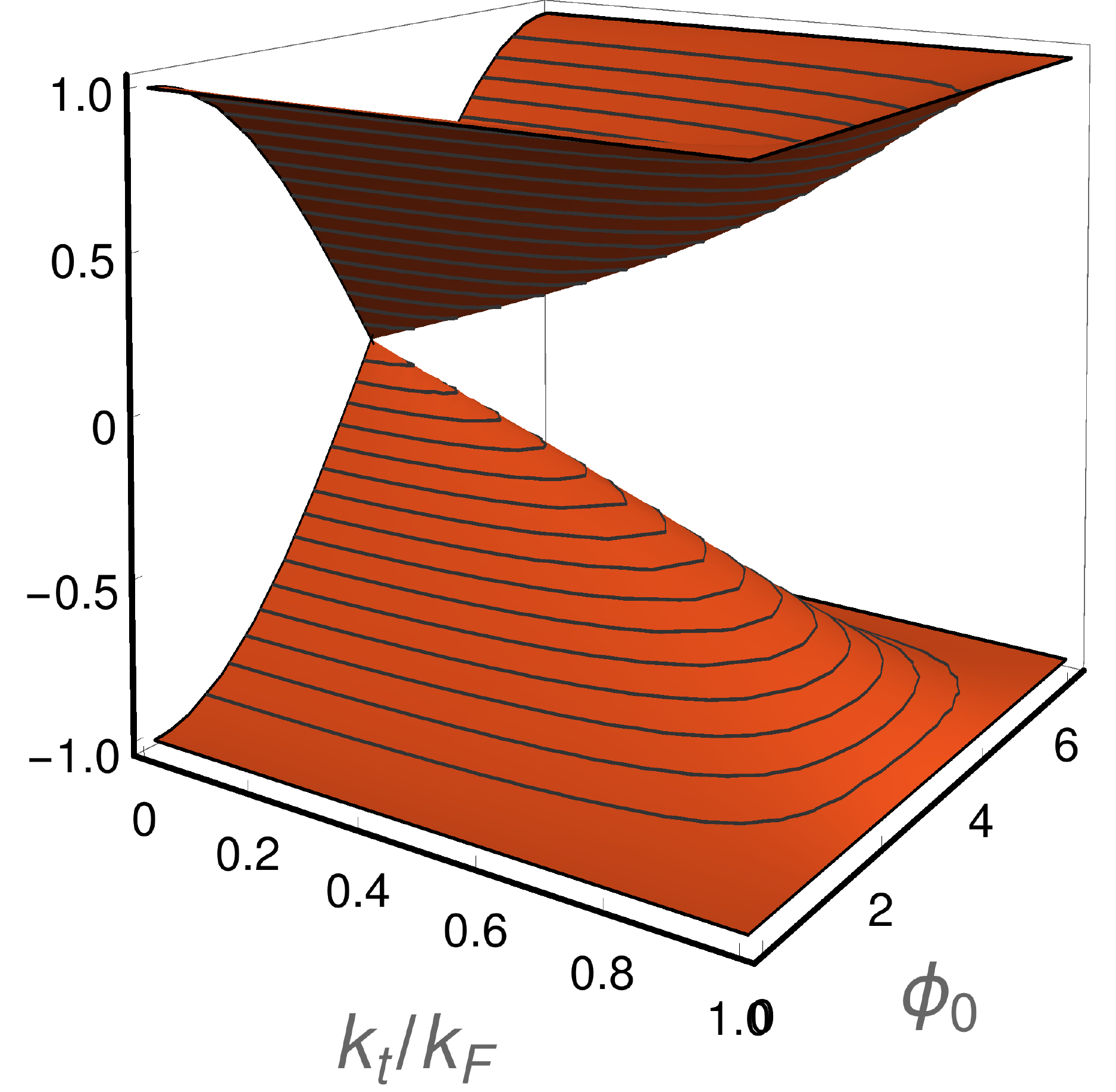}}
{\includegraphics*[width=0.48\linewidth]{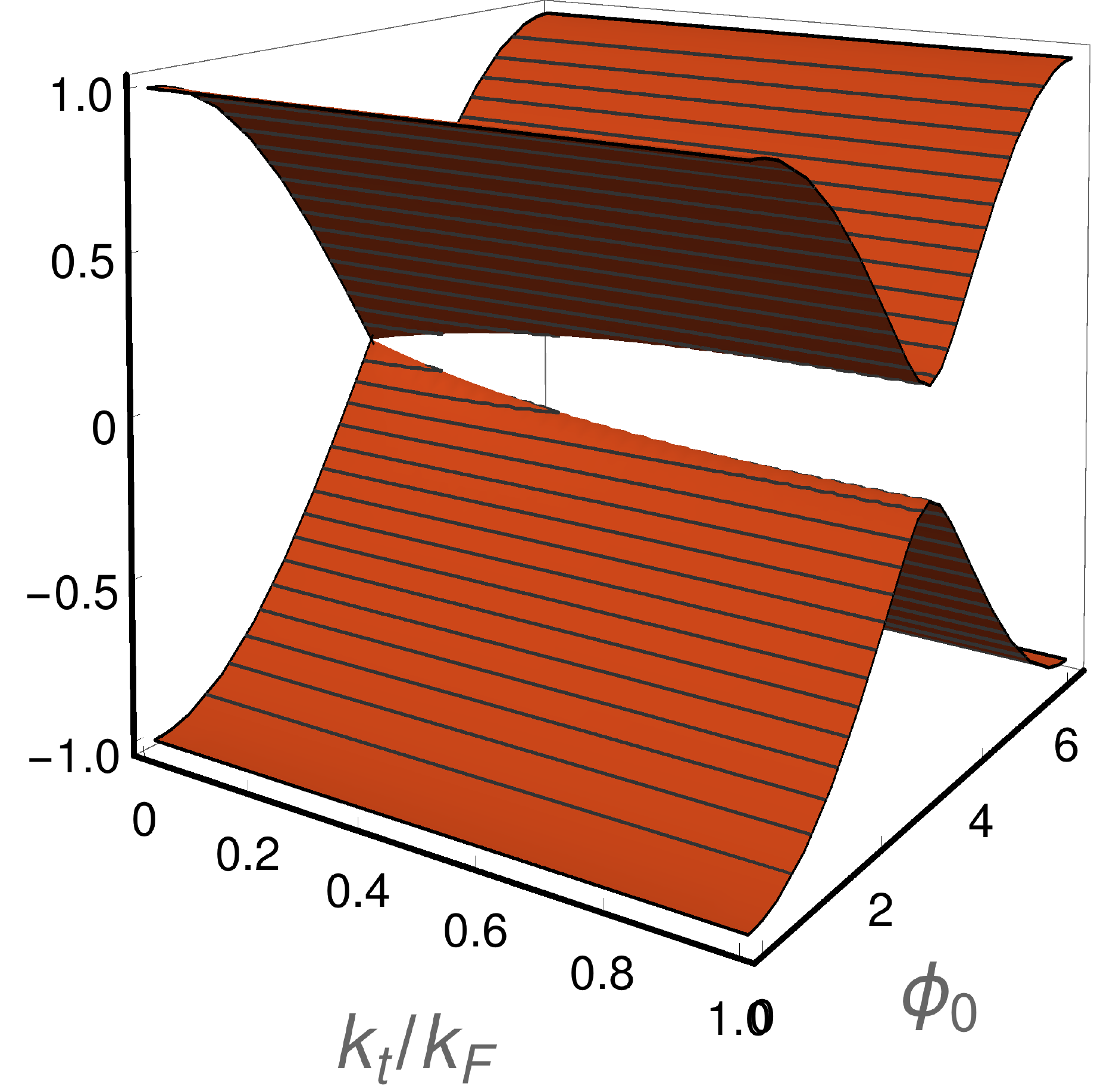}} \caption{Plot
of the Andreev bound sates $E_{\pm}/\Delta_0$ as a function of
$k_t/k_F$ and $\phi_0$ for $\chi= \phi_{\vec k}=\pi/4$. The left
panel corresponds to $n_1=n_2=1$ and the right panel to $n_1=1$ and
$n_2=2$. For both panels $\mu_0=\mu'_0=100\Delta_0$ and all energies
are scaled in units of $\Delta_0$. See text for details.}
\label{fig2}
\end{figure}

To obtain the Andreev bound states, we now impose the usual current
continuity condition along $\hat z$. For Weyl or multi-Weyl
electrons with linear longitudinal dispersion, it is well known that
current continuity amounts to continuity of the wavefunction leading
to the conditions
\begin{eqnarray}
\psi_I(z=0) &=& \psi_{II}(z=0),  \quad \psi_{II}(z=d) =
\psi_{III}(z=d) \nonumber\\ \label{bcond1}
\end{eqnarray}
Eq.\ \ref{bcond1} leads to eight linear homogeneous equations. The
energy of the Andreev bound states is to be found by demanding
non-zero solutions of these equations \cite{ds1,ks2,been2}. Here we
concentrate on the regime $\mu_0, \mu'_0 \gg \Delta_0$, for which
$\theta_{\vec k (2)[(4)]}=-\theta_{\vec k (1)[(3)]}$. In this
regime, it is possible to simplify these equations significantly.
Eliminating $p$, $q$, $r$ and $s$ from these equations in this
regime, one gets, after a straightforward calculation, a set of four
linear homogeneous equations involving $a_{1,2}$ and $b_{1,2}$.
These equations are given by
\begin{widetext}
\begin{eqnarray}
&&a_1 \left(\cos\theta_{\vec k 2} \sin(\theta_{\vec k
1}-\theta_{\vec k 2}) e^{i (\gamma + k_z d)} + \sin \theta_{\vec k
2} \cos (\theta_{\vec k 1}+\theta_{\vec k 2}) e^{i (\gamma - k_z d)}
\right) + b_1  \left(\cos\theta_{\vec k 2} \cos (\theta_{\vec k
1}-\theta_{\vec k 2}) e^{i k_z d} \right. \nonumber\\
&& \left.- \sin \theta_{\vec k 2} \sin (\theta_{\vec k
1}+\theta_{\vec k 2}) e^{-i k_z d} \right) = \cos(2 \theta_{\vec k
2}) e^{i(n_1-n_2)\phi_{\vec k}} \left( a_2 \cos \theta_{\vec k 3}
e^{i (\gamma + k_z^{s(3)} d)} - b_2 \sin \theta_{\vec k 3} e^{i
(\phi_0 - k_{z}^{s(4)}) d} \right) \nonumber\\
&&a_1 \left(\sin\theta_{\vec k 2} \sin(\theta_{\vec k
1}-\theta_{\vec k 2}) e^{i (\gamma + k_z d)} + \cos \theta_{\vec k
2} \cos (\theta_{\vec k 1}+\theta_{\vec k 2}) e^{i (\gamma - k_z d)}
\right) + b_1  \left(\sin \theta_{\vec k 2} \cos (\theta_{\vec k
1}-\theta_{\vec k 2}) e^{i k_z d} \right. \nonumber\\
&& \left.- \cos \theta_{\vec k 2} \sin (\theta_{\vec k
1}+\theta_{\vec k 2}) e^{-i k_z d} \right) = \cos(2 \theta_{\vec k
2}) e^{-i(n_1-n_2)\phi_{\vec k}} \left( a_2 \sin \theta_{\vec k 3}
e^{i (\gamma + k_z^{s(3)} d)} - b_2 \cos \theta_{\vec k 3} e^{i
(\phi_0 - k_{z}^{s(4)}) d} \right)\nonumber\\
&&a_1 \left(\cos\theta'_{\vec k 2} \sin(\theta_{\vec k
1}+\theta'_{\vec k 2}) e^{i k'_z d} - \sin \theta'_{\vec k 2} \cos
(\theta_{\vec k 1}-\theta'_{\vec k 2}) e^{- ik'_z d)} \right) + b_1
\left(\cos\theta'_{\vec k 2} \cos (\theta_{\vec k
1}+\theta'_{\vec k 2}) e^{i (\gamma-k'_z d)} \right. \nonumber\\
&& \left.+ \sin \theta'_{\vec k 2} \sin (\theta_{\vec k
1}-\theta'_{\vec k 2}) e^{i (\gamma+k_z d)} \right) = \cos(2
\theta'_{\vec k 2}) e^{i(n_1-n_2)\phi_{\vec k}} \left( a_2 \cos
\theta_{\vec k 3} e^{-i (\phi_0- k_z^{s(3)} d)} - b_2 \sin
\theta_{\vec k 3} e^{i
(\gamma - k_{z}^{s(4)}) d} \right)\nonumber\\
&&a_1 \left(\cos\theta'_{\vec k 2} \sin(\theta_{\vec k
1}+\theta'_{\vec k 2}) e^{i k'_z d} + \sin \theta'_{\vec k 2} \cos
(\theta_{\vec k 1}-\theta'_{\vec k 2}) e^{- ik'_z d)} \right) + b_1
\left(\cos \theta'_{\vec k 2} \sin (\theta'_{\vec k
2}-\theta_{\vec k 1}) e^{i (\gamma -k'_z d)} \right. \nonumber\\
&& \left.- \sin \theta'_{\vec k 2} \cos (\theta_{\vec k
1}+\theta'_{\vec k 2}) e^{-i (k'_z d -\gamma)} \right) = \cos(2
\theta'_{\vec k 2}) e^{-i(n_1-n_2)\phi_{\vec k}} \left( a_2 \sin
\theta_{\vec k 3} e^{-i (\phi_0- k_z^{s(3)} d)} + b_2 \cos
\theta_{\vec k 4} e^{i (\gamma - k_{z}^{s(4)}) d} \right)
\label{thickb}
\end{eqnarray}
\end{widetext}
In what follows, we shall numerically solve Eq.\ \ref{thickb} to
obtain the expression for Andreev bound states while discussing the
properties of barriers away from the thin barrier limit. In the rest
of this section, we concentrate on the thin barrier limit. In this
limit Eqs.\ \ref{thickb} can be further simplified to yield analytic
expression for the Andreev bound states. For this purpose, we first
note that in the thin barrier limit $k_z d , k'_z d \to \chi$ and
$k_{z}^{s (3)} d , k_{z}^{s (4)} d \to 0$. Moreover, in this limit
$\theta_1$ and $\theta_3$ becomes independent of $\zeta$: $\theta_1
[\theta_3] \to \arcsin (\epsilon_0 |\vec
k_t|^{n_1[n_2]}/\mu_0[\mu'_0])/2$. Defining $\alpha_0 =
(n_1-n_2)\phi_{\vec k}/2 + \chi$, we find that in the thin barrier
limit Eq.\ \ref{thickb} simplifies to ${\mathcal N} A_t =0$, where
$A_t= (a_1,b_1,a_2,b_2)^T$ is a four-component column vector and the
matrix ${\mathcal N}$ is given by
\begin{widetext}
\begin{eqnarray}
{\mathcal N} &=& \left( \begin{array}{cccc} \sin(\theta_{\vec k 1})
e^{i\gamma} & \cos(\theta_{\vec k 1}) & -\cos(\theta_{\vec k 3})
e^{i (\alpha_0 + \gamma)} & \sin(\theta_{\vec k 3}) e^{i(\alpha_0 +
\phi_0)} \\ \cos(\theta_{\vec k 1}) e^{i\gamma} & -\sin(\theta_{\vec
k 1} & -\sin(\theta_{\vec k 3}) e^{i (-\alpha_0 + \gamma)} &
-\cos(\theta_{\vec k 3}) e^{i(-\alpha_0 + \phi_0)} \\
\sin(\theta_{\vec k 1}) & \cos(\theta_{\vec k 1}e^{i \gamma} &
-\cos(\theta_{\vec k 3}) e^{i (\alpha_0 -\phi_0)} &
\sin(\theta_{\vec k 3}) e^{i(\alpha_0 + \gamma)} \\
\cos(\theta_{\vec k 1}) & -\sin(\theta_{\vec k 1} e^{i\gamma} &
-\sin(\theta_{\vec k 3}) e^{-i (\alpha_0 + \phi_0)} &
-\cos(\theta_{\vec k 3}) e^{i(\gamma-\alpha_0)} \end{array} \right)
\label{thinb}
\end{eqnarray}
\end{widetext}
Note that the dependence of the matrix elements of ${\mathcal N}$ on
$\chi$ comes only through $\alpha_0$ in the thin barrier limit. To
obtain the analytic expression of the Andreev bound states, we
demand ${\rm Det}[{\mathcal N}]=0$, which yields the dispersions of
these states in terms of the normal state transmission $T_N(\vec
k_t) \equiv T_N$ of the junction \cite{ds1} as
\begin{widetext}
\begin{eqnarray}
E_{\pm}(\phi_0) &=& \pm \Delta_0 \sqrt{1- T_N \sin^2(\phi_0/2)},
\quad T_N = \frac{\cos(2 \theta_{\vec  k 1}) \cos (2 \theta_{\vec k
3})}{(\cos \theta_{\vec k 1} \cos \theta_{\vec k 3} +\sin
\theta_{\vec k 1} \sin \theta_{\vec k 3})^2 - \sin(2 \theta_{\vec k
1}) \sin(2 \theta_{\vec k 3}) \cos^2 \alpha_0} \label{andreev1}
\end{eqnarray}
\end{widetext}
A plot of $E_{\pm}/\Delta_0$ as a function of the relative phase
$\phi_0$ and the transverse momentum $|\vec k_t|\equiv k_t$ is shown
in Fig.\ \ref{fig2} in the thin barrier limit with $\chi= \phi_{\vec
k}=\pi/4$ for $n_1=1$ and $n_2=1(2)$ in the left(right) panels. For
these plots, we find that $E_{+}$ and $E_-$ touches $\phi_0=\pi/2$
for $k_t=0$, which satisfies $T_N(k_t=0)=1$ due to Klein tunneling
\cite{ks1,been1}. The behavior of the bound state spectrum near this
touching point depends crucially on whether the topological winding
number changes across the junction; for $n_1=n_2=1$ (left panel),
the spectrum is isotropic around $k_t=0$ with a large slope, while
for $n_2=2n_1=2$ (right panel), the spectrum around $k_t=0$ has much
lower slope.

Using Eq.\ \ref{andreev1}, one can find the expression of the
Josephson current through the junction. Since the dispersing states
with $E>\Delta_0$ do not depend on $\phi_0$, the Josephson current
in the system at a temperature $T_0 \ll \Delta_0/k_B$ (where $k_B$
is the Boltzaman constant) is determined solely by the bound states
and is given by \cite{ks2,been2}
\begin{eqnarray}
I_J &=& \frac{e \Delta_0}{2 \hbar} \sum_{\vec k_t} \frac{T_N
\sin(\phi_0)}{\sqrt{1-T_N \sin^2 (\phi_0/2)}} \tanh
\left[\frac{E_+(\phi_0)}{2 k_B T_0} \right] \nonumber\\
\label{jjcurrent}
\end{eqnarray}

Using Eq.\ \ref{jjcurrent}, we now demonstrate the barrier
independence of $I_c R_N$. To this end, we note that the sum over
transverse momenta in Eq.\ \ref{jjcurrent} can be replaced by
integral over $(k_x,k_y)$ for large enough sample. To carry out this
integral, we use the parametrization
\begin{eqnarray}
k_x [k_y] &=& k_F \left[\sin 2 \theta \right]^{1/n_1} \cos \phi
[\sin \phi ]
\nonumber\\
k_z &=&  k_F \cos 2\theta  \label{trans1}
\end{eqnarray}
One can equivalently parametrize $\vec k_t$ using $n_2$ and $\mu'_0
= \hbar v_F k'_F$, where $k'_F $ is the Fermi wave vector in region
III. In what follows we shall always choose the parametrization with
lower $n$ (chosen to be $n_1$ without loss of generality in the rest
of this work) for carrying out the integrals; this ensures that the
integral over $\theta$ has the range $-\pi/4 \le \theta \le \pi/4$.
Using this parametrization one gets
\begin{eqnarray}
I_J &=& \frac{I_0}{4n_1} \left(\frac{L k_F}{2 \pi}\right)^2
\int_0^{\pi/4} d\theta \int_0^{2 \pi} d \phi \nonumber\\
&& \times (\sin 2 \theta)^{2/n_1-1} \cos 2 \theta \frac{T_N \sin
\phi_0}{\sqrt{1- T_N \sin^2 (\phi_0/2) }}, \label{jjcurrent1}
\end{eqnarray}
where $L$ is the transverse dimension of the junction.

\begin{figure}
\rotatebox{0}{\includegraphics*[width=0.48\linewidth]{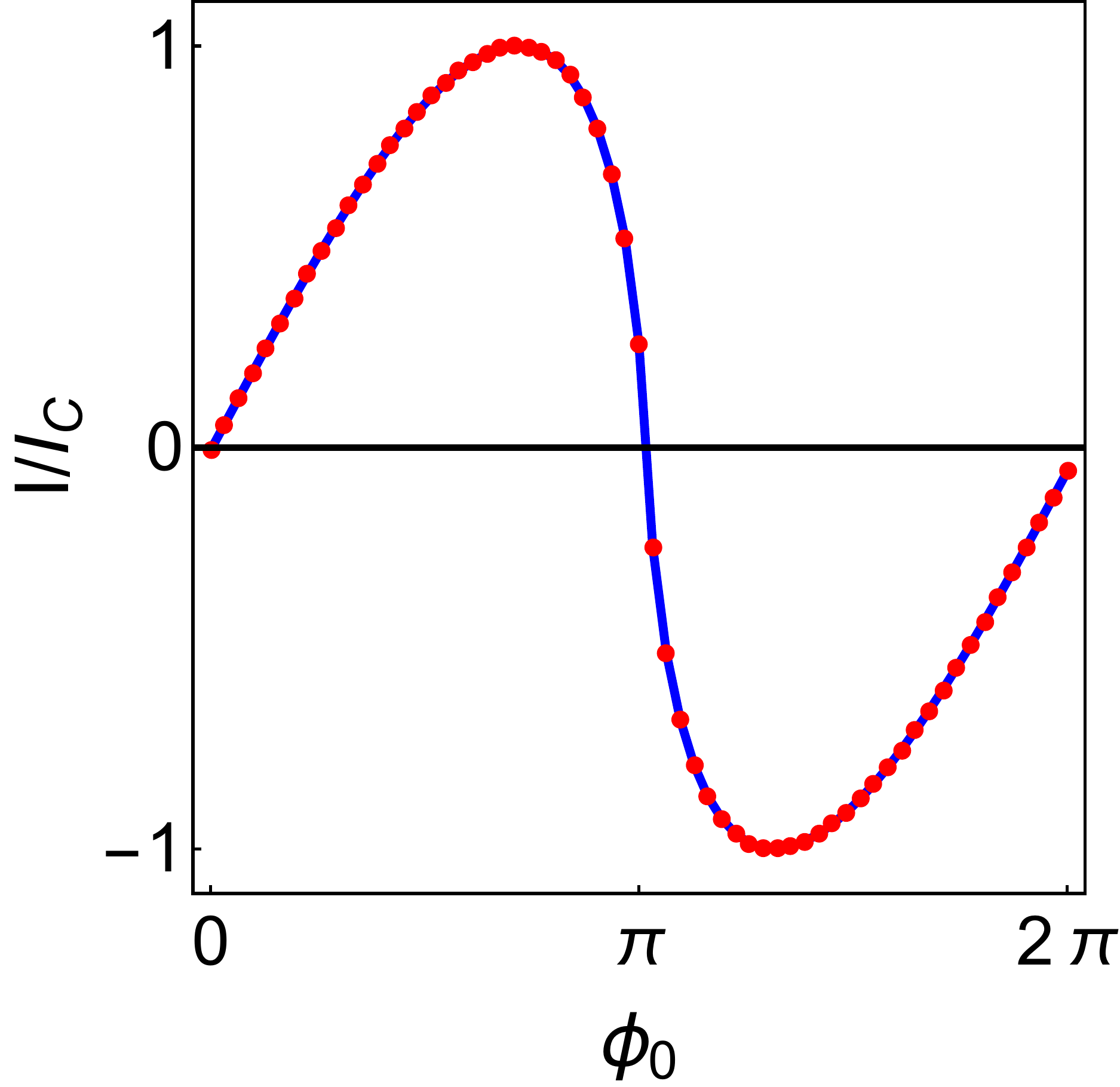}}
{\includegraphics*[width=0.48 \linewidth]{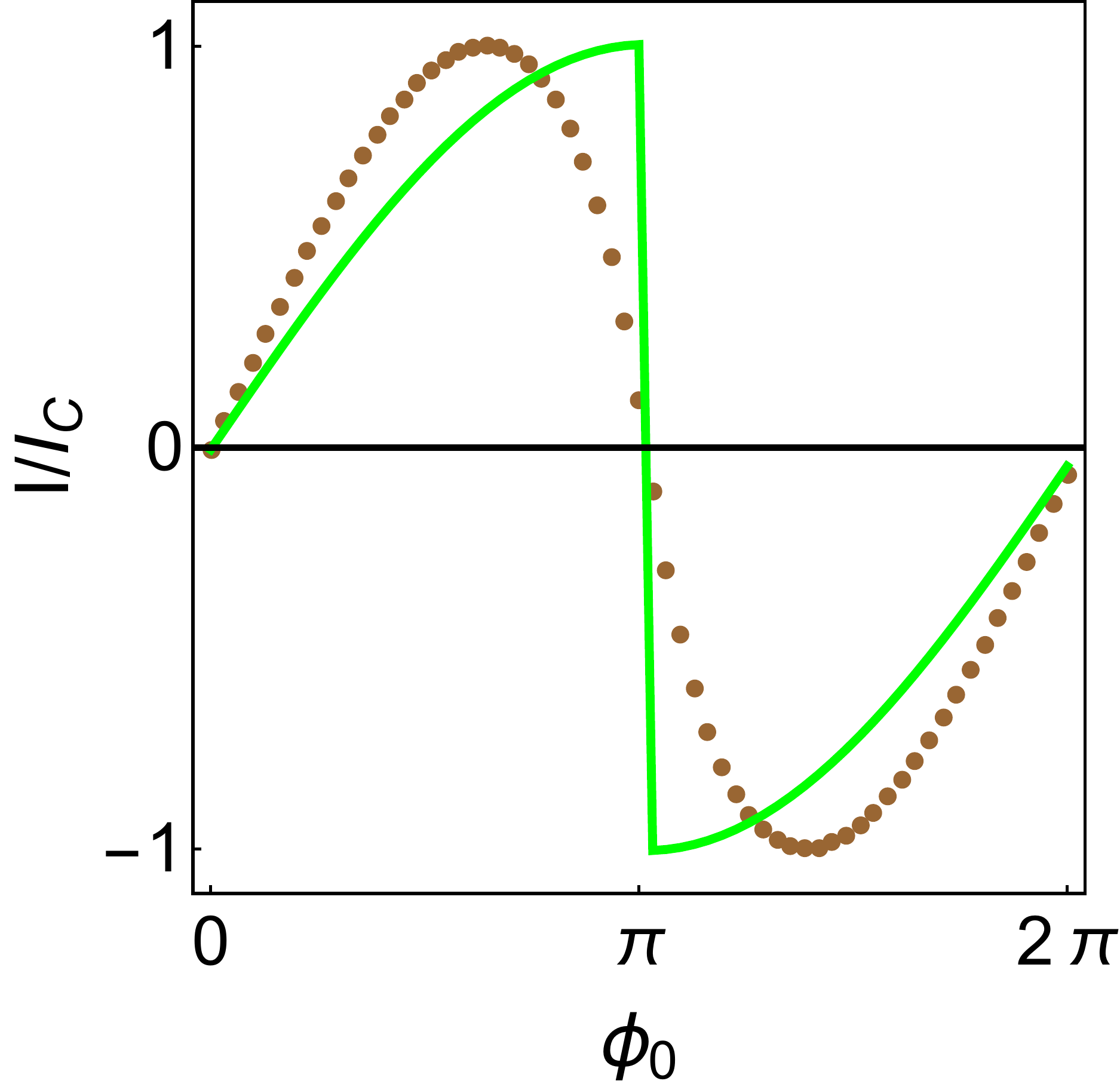}} \caption{ Plot
of $I_J/I_c$ as a function of $\phi_0$ for $\chi=\pi$ (blue [green]
solid lines in left[right] panels) and $\chi=\pi/4$ (red dots) with
$n_1=1$ and $n_2=2$ (left panel) and $n_1=n_2=1$ (right panel). Note
that $I_J/I_c$ becomes independent of $\chi$ in the thin barrier
limit for $n_1 \ne n_2$ (left panel) but depends substantially on
$\chi$ if $n_1=n_2$ (right panel). All other parameters are same as
in Fig.\ \ref{fig2}. See text for details. } \label{fig3}
\end{figure}

Next, we denote that value of $\phi_0$ for which $I_J$ is maximum to
be $\phi_0^{m}$. This maximum is obtained by demanding that
$\phi_0^m$ is a solution of $\partial_{\phi_0} I_J=0$ and is given
by
\begin{eqnarray}
\phi_0^{m}= \arccos[[(T_N-2) + 2 \sqrt{1-T_N}]/T_N]
\label{phimaxsol}
\end{eqnarray}
Note that $\phi_0^m$ depends on $\vec k_t$ through $T_N$. Using
Eqs.\ \ref{trans1} and \ref{phimaxsol} and defining $I_0 =
2e\Delta_0/\hbar$ one finds, at $T_0=0$,
\begin{eqnarray}
I_c &=& I_J[\phi_0=\phi_0^m] = \frac{I_0}{n_1} \left(\frac{L k_F}{2
\pi}\right)^2 {\mathcal I}_1 \nonumber\\
{\mathcal I_1} &=& \frac{1}{4} \int_0^{\pi/4} d\theta \int_0^{2 \pi}
d \phi (\sin 2\theta)^{2/n_1-1} \cos 2\theta \nonumber\\
&& \times \frac{T_N \sin \phi_0^m}{\sqrt{1- T_N \sin^2 (\phi_0^m/2)
}}\label{icexp}
\end{eqnarray}
A plot of $I_J/I_c$ as a function of $\phi_0$ is shown in the left
panel of Fig.\ \ref{fig3} for two different values of $\chi$ and
$n_1=1$ and $n_2=2$. We find that $I_J/I_c$ is independent of
$\chi$; this feature is a consequence of integration over the
azimuthal angle $\phi_{\vec k}$ which eliminates the $\chi$
dependence, provided $n_1 \ne n_2$ (Eqs.\ \ref{jjcurrent} and
\ref{icexp}). In contrast, for $n_1=n_2$, varying $\chi$ leads to
substantial change in $I_J/I_c$ as shown in the right panel of Fig.\
\ref{fig3}. Thus we find a qualitative difference between dependence
of $I_J$ on $\chi$ for junctions , where the topological winding
number changes across the junction and for those , where it does not.

For computing $R_N$, one first finds the normal state junction
conductance $G_N = e^2/h \sum_{\vec k_t} T_N $. Using this, one
finds the normal state resistance $R_N = G_N^{-1}$, with $G_N$ given
by
\begin{eqnarray}
G_N &=& \left( \frac{e^2}{h n_1} \frac{L k_F}{2 \pi}\right)^2
\int_0^{\pi/4} d\theta \int_0^{2 \pi} d \phi \nonumber\\
&& \times (\sin 2 \theta)^{2/n_1-1} \cos 2\theta \, \, T_N  \nonumber\\
&=& \frac{e^2}{2 n_1 \pi \hbar} \left(\frac{L k_F}{2 \pi}\right)^2
{\mathcal I}_2 \label{rnexp}
\end{eqnarray}
From the expression of $T_N$ in Eq.\ \ref{andreev1}, it is evident
that $I_N$ depends on $\phi$ and $\chi$ only through $\alpha_0$.
Consequently, it is possible to carry out the integral over $\phi$
analytically; in particular, it is easy to show by expressing the
integral over $\phi$ as a contour integral with the substitution $z=
\exp[i \phi]$ \cite{ds1}, that $\int^{2 \pi}_0 d \phi T_N$ is
independent of $\chi$ for $n_1 \ne n_2$. A similar $\chi$
independence can be shown for ${\mathcal I}_1$ (Eq.\ \ref{icexp})
required to evaluate $I_c$. Note that for $n_1 = n_2$, $T_N$ becomes
independent of $\phi$ and the $\chi$ dependence is retained. Thus we
find that the change in topological winding number across the
junction is crucial for both $I_c$ and $R_N$ to be independent of
the barrier strength in the thin barrier limit. Using Eqs,
\ref{icexp} and \ref{rnexp}, we find
\begin{eqnarray}
I_c R_N  &=& \frac{\pi \Delta_0}{2 e} \frac{{\mathcal
I}_1}{{\mathcal I}_2} = \frac{\pi \Delta_0}{2 e} c,\label{icrnexp}
\end{eqnarray}
where the ratio $c={\mathcal I}_1/{\mathcal I}_2$ depends on $n_1$
and $n_2$.

\begin{figure}
\rotatebox{0}{\includegraphics*[width=0.48 \linewidth]{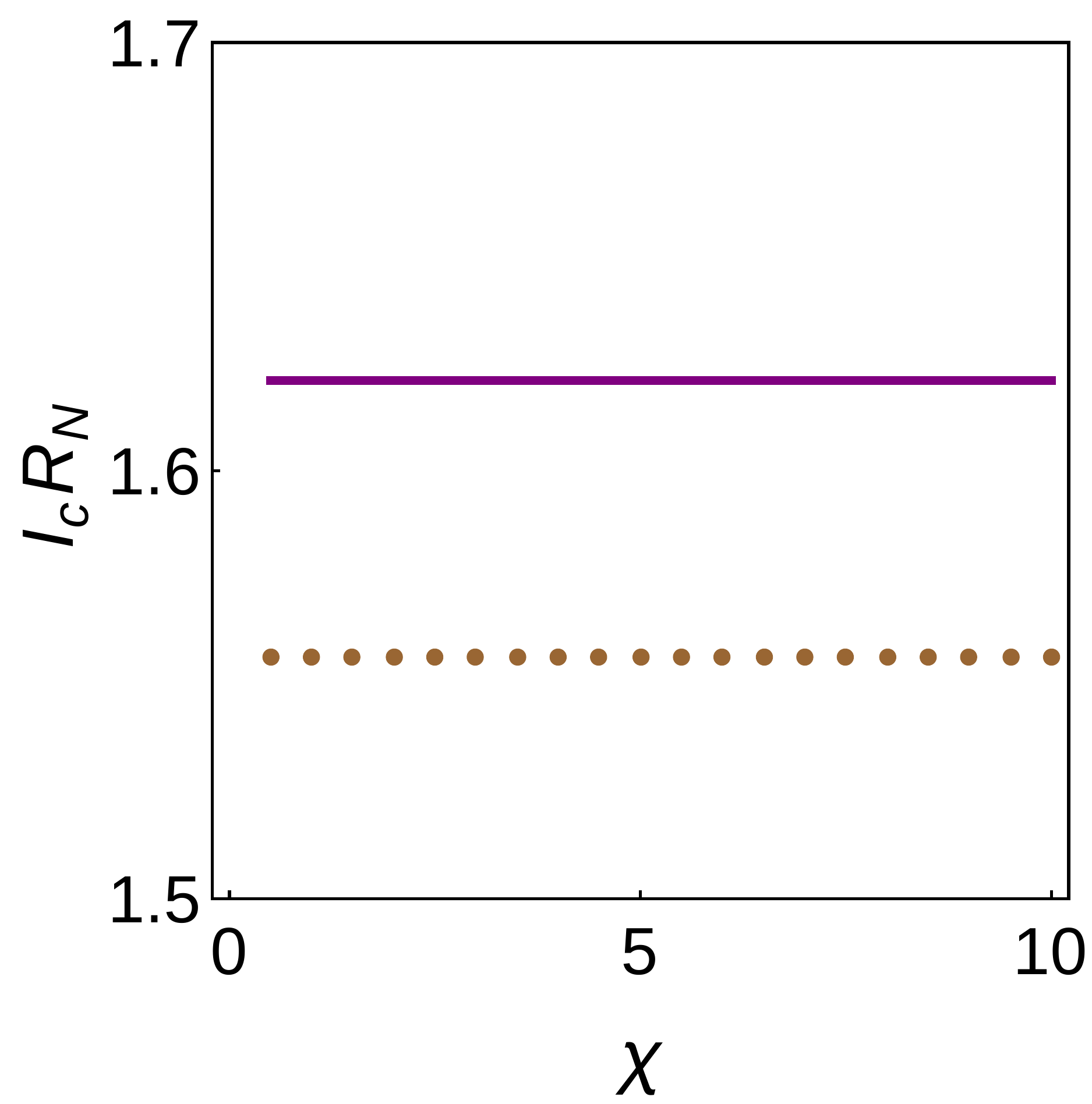}}
{\includegraphics*[width=0.48 \linewidth]{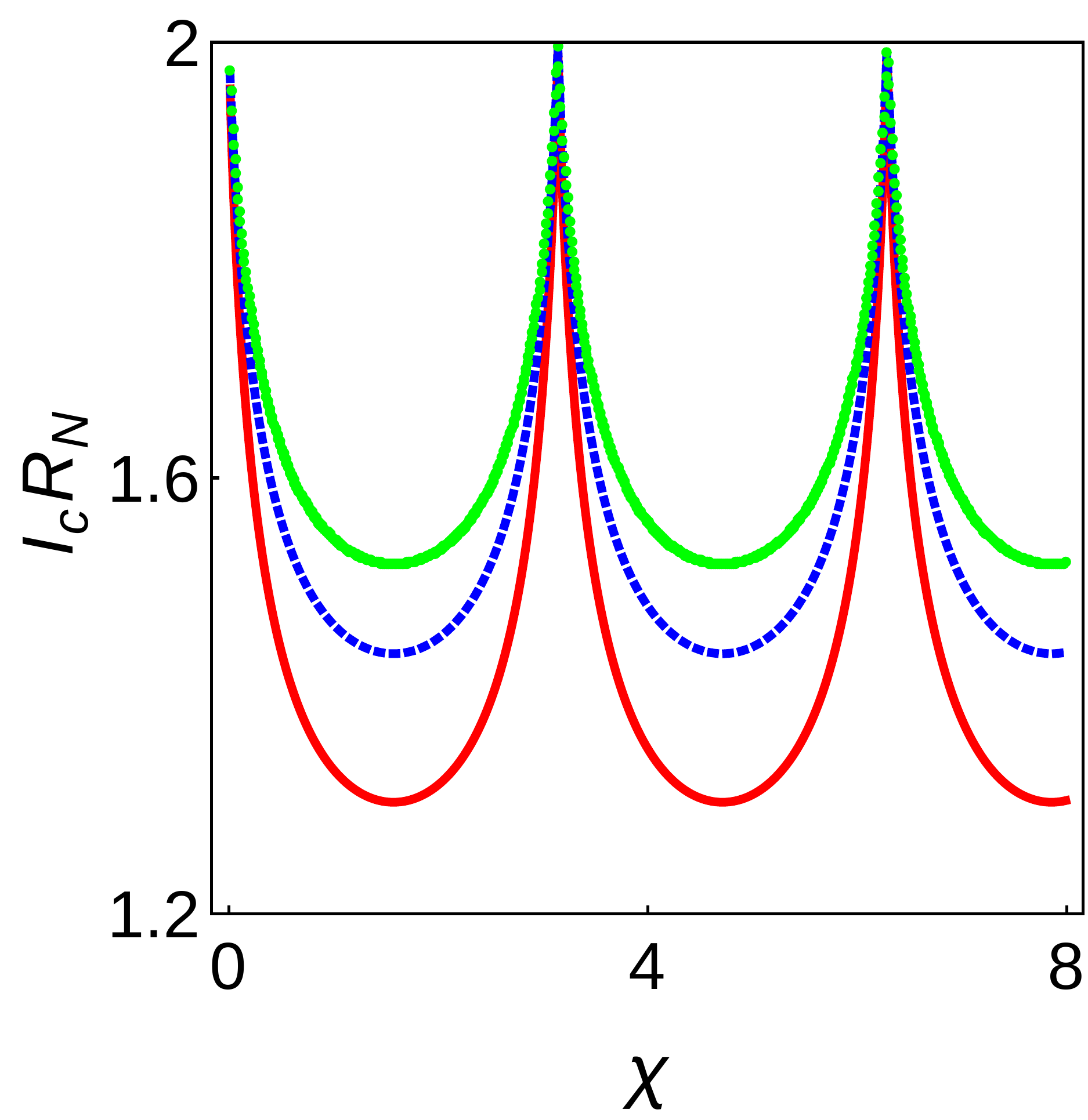}} \caption{ Plot
of $I_c R_N$ in units of $\pi \Delta_0/(2e)$ as a function of
$\chi$. The left panel represents cases for $n_1\ne n_2$; the brown
dotted (violet solid) line represents $n_1=1(2)$ and $n_2=2(3)$. The
right panel represents case for which $n_1=n_2$; the red solid, blue
dashed and the green dotted lines corresponds to $n_1=n_2=1$, $2$
and $3$ respectively. All other parameters are same as in Fig.\
\ref{fig2}. See text for details. } \label{fig4}
\end{figure}

A plot of $I_c R_N$ as a function of $\chi$ is shown in Fig.\
\ref{fig4}. The left panel of Fig.\ \ref{fig4} shows the barrier
independent of $I_c R_N$ for $n_1 \ne n_2$ for several choice of
$n_1$ and $n_2$. This allows us to compute $c$; we find $c=1.56
(1.62)$ for $n_1=1$ and $n_2=2(3)$. This demonstrates the weak
dependence of $c$ on $n_1$ and $n_2$. In contrast, $c$ is an
oscillatory function of $\chi$ for $n_1=n_2$ as shown in right panel
of Fig.\ \ref{fig4}. These plots therefore demonstrates that $I_c
R_N$ in these junctions becomes a barrier independent universal
constant for a fixed $n_1 \ne n_2$; the value of this constant
depends weakly on $n_1$ and $n_2$. This behavior is in sharp
contrast to all JJs made out of topological or conventional
superconductors studied earlier \cite{ko1,ab1,ks2,been2}.

\begin{figure}
\rotatebox{0}{\includegraphics*[width=0.98 \linewidth]{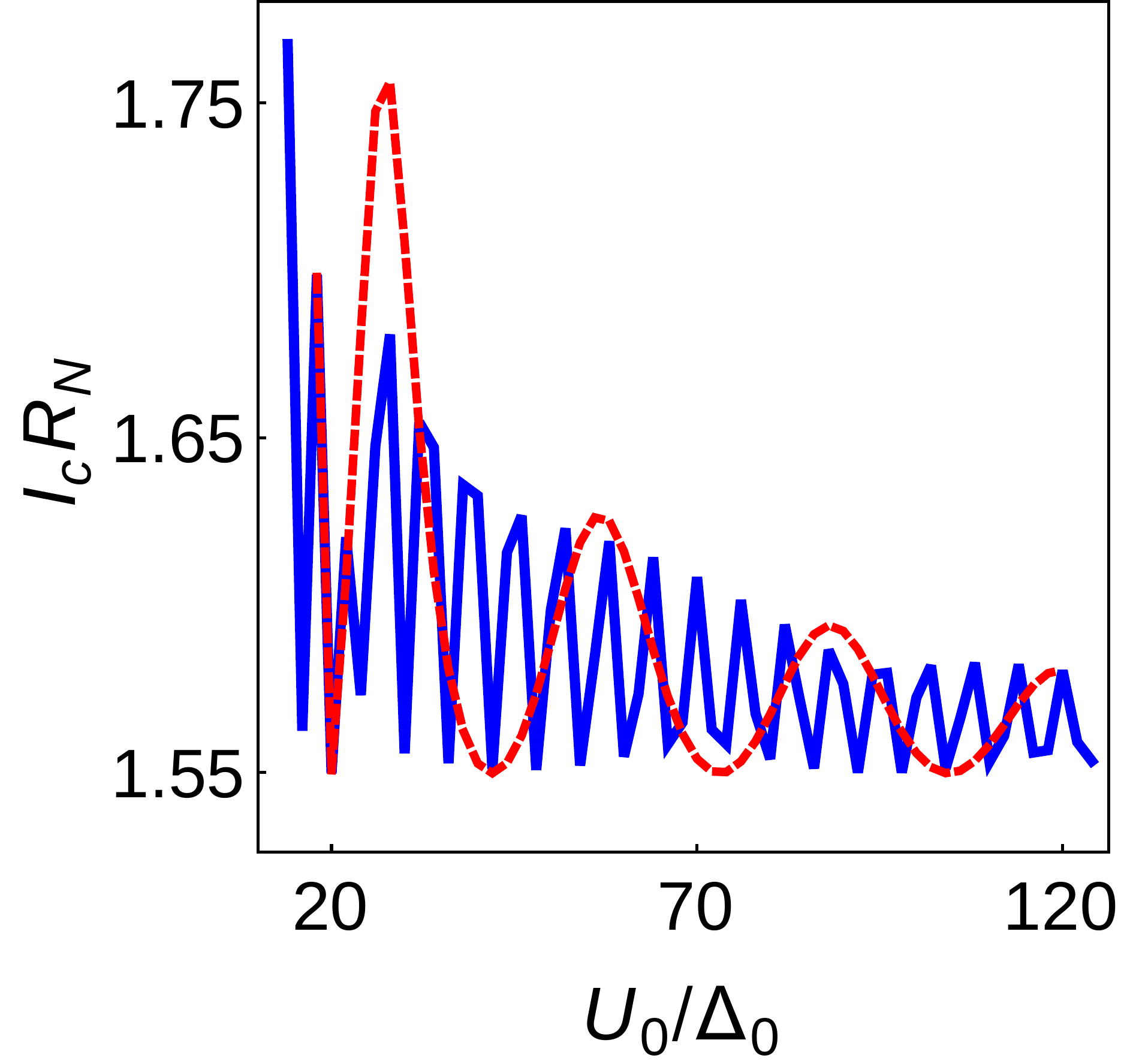}}
\caption{ Plot of $I_c R_N$ in units of $\pi \Delta_0/(2e)$ as a
function of $U_0/\Delta_0$ for $k_F d=0.5$ (blue solid line) and
$0.1$ (red dotted line). Here all energies are scaled in units of
$\Delta_0$, and $\mu_0=\mu'_0=10 \Delta_0$. See text for details. }
\label{fig5}
\end{figure}

Next, we move away from the thin barrier limit and allow for
arbitrary width $d$ of the barrier potential. To this end, we
numerically solve Eq.\ \ref{thickb} for $\mu_0= \mu'_0 \gg \Delta_0$
for a given transverse momentum $(k_x, k_y)$ and find the Andreev
bound state energy as a function of $\phi_0$. Next, we follow Eqs.\
\ref{jjcurrent1}, \ref{icexp}, and \ref{rnexp} to compute $I_c R_N$
numerically. The result of this computation is shown in Fig.\
\ref{fig5} for $2n_1=n_2=2$. We find that with increasing barrier
thickness, $I_c R_N$ starts to oscillate with the barrier strength
$U_0/\Delta_0$. However, the magnitude of this oscillation decreases
rapidly with decreasing $d$ as can be seen by comparing the plots
for $k_0 d=0.1$ and $k_0 d=0.5$ in Fig.\ \ref{fig6}. Moreover, the
amplitude of these oscillations, for $k_0 d \simeq 1$ is found to be
small compared to the JJs involving Weyl semimetals with
$n_1=n_2=1$, particularly for $U_0 \gg \Delta_0$. These oscillations
can thus be distinguished from the ones arising in JJs involving
Weyl semimetals with $n_1=n_2$. The amplitude of these oscillations
in JJs with $n_1 \ne n_2$ for a fixed $d$ is a monotonically
decaying function of $U_0/\Delta_0$, while those in JJs with
$n_1=n_2$ is almost independent of the barrier height for large
$U_0/\Delta_0$.

\section{AC Josephson effect}
\label{acj}

\begin{figure}
\rotatebox{0}{\includegraphics*[width=0.48 \linewidth]{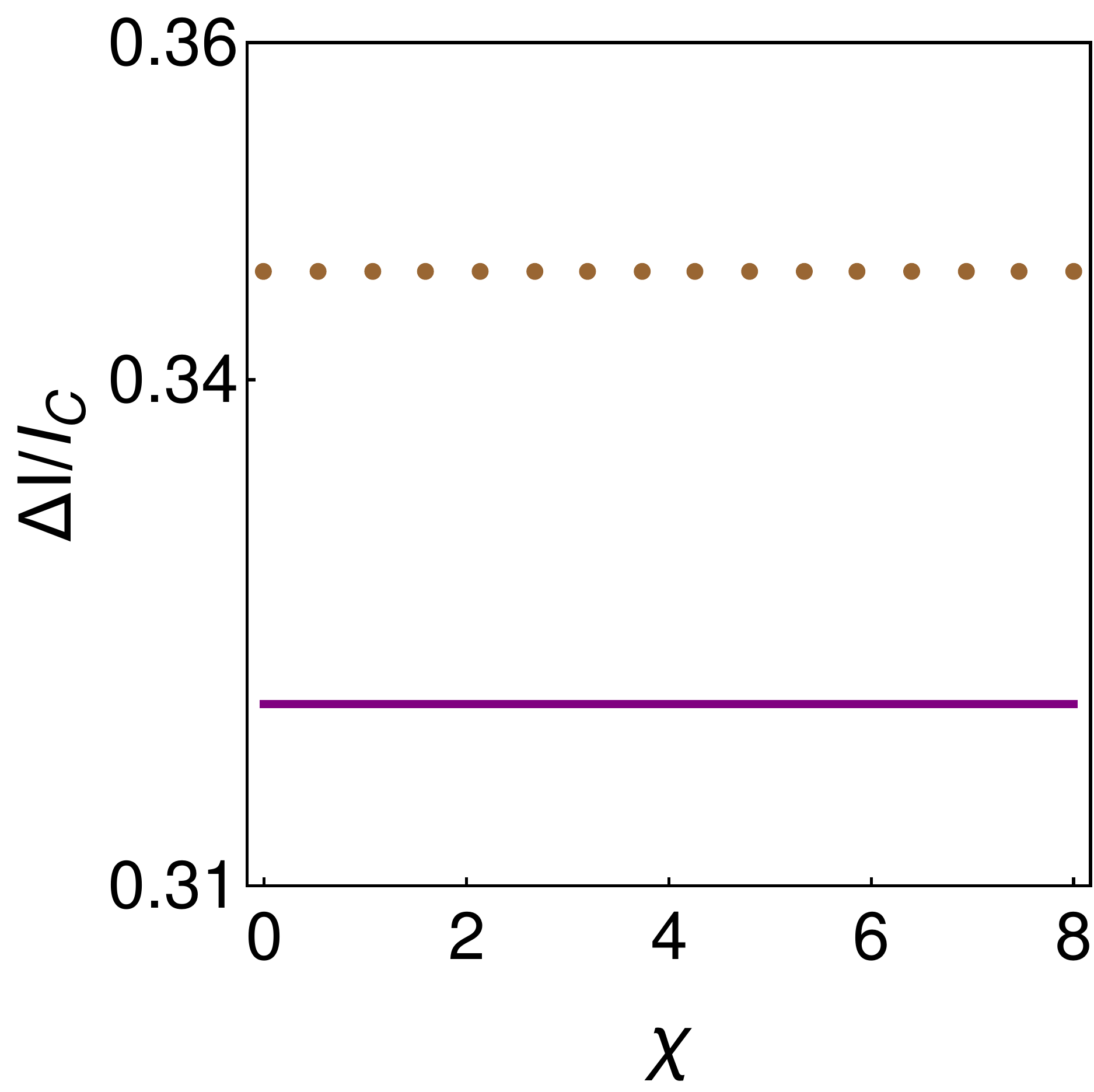}}
{\includegraphics*[width=0.48 \linewidth]{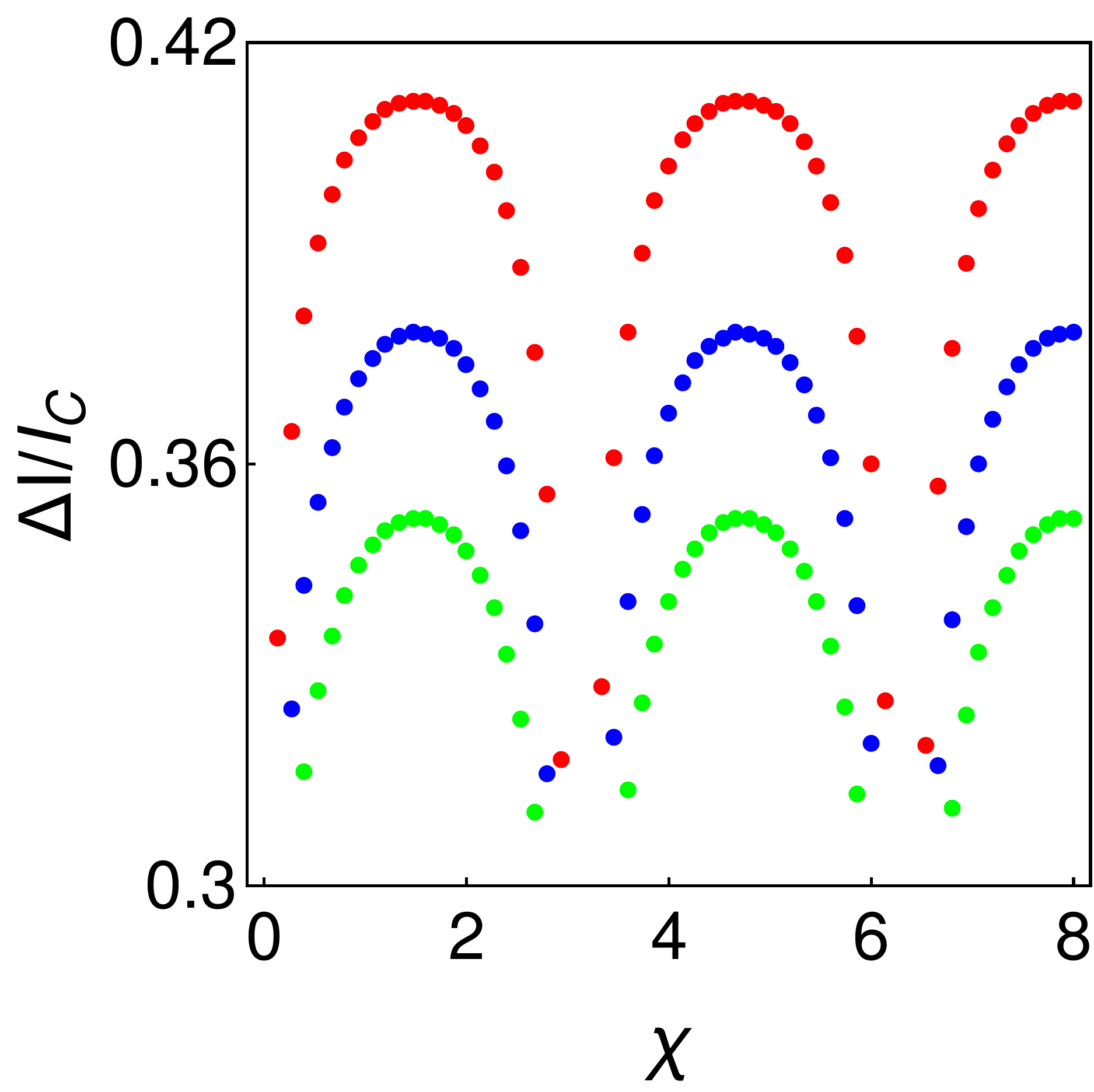}} \caption{ Plot
of the Shapiro step width $\Delta I/I_c \equiv (\Delta
I)_{m_0=1}/I_c$ as a function of $\chi$ for $m_0=1$ and $V_1/(\hbar
\omega_D)=1$. The left panel shows $\Delta I/I_c$ for $n_1=1$ and
$n_2=2$ (brown dotted line) and $n_2=3$ (purple solid line). The
right panel corresponds to $n_1=n_2$ with $n_1=1$ (red dotted line)
$n_1=2$ (blue dotted line) and $n_1=3$ (green dotted line). All
other parameters are same as in Fig.\ \ref{fig2}. See text for
details. } \label{fig6}
\end{figure}

In this section we analyze the AC Josephson effect in topological
junctions, where the topological winding number changes across the
junction. To this end, we first consider the voltage biased
junctions which are analytically more tractable compared to their
current biased counterparts. For such junctions, the expressions of
the Josephson current, in the presence of bias voltage and a
microwave radiation, may be obtained by the usual substitution
\begin{eqnarray}
\phi_0 &\to& \phi_0 + \frac{2e}{\hbar} \int^t dt' [V_0 + V_1
\cos(\omega_D t')], \label{vbias1}
\end{eqnarray}
where $V_0[V_1]$ are the amplitude of the DC voltage [microwave
radiation] and $\omega_D$ is the frequency of radiation.
Substituting this in Eq.\ \ref{jjcurrent} and using the identity
$\exp[i a \sin(\omega_D t)] = \sum_{m=-\infty}^{\infty} J_m(a)
\exp[i m \omega_D t]$, where $J_m$ denotes $m^{\rm th}$ order Bessel
function and $m$ takes integer values, we find
\begin{widetext}
\begin{eqnarray}
I_J(t) &=& \frac{I_0}{4n_1} \left(\frac{L k_F}{2 \pi}\right)^2
\int_0^{\pi/4} d\theta \int_0^{2 \pi} d \phi \frac{T_N
\sum_{m=-\infty}^{\infty} (\sin 2\theta)^{2/n_1-1} \cos 2 \theta
J_m\left(\frac{2eV_1}{\hbar \omega_D}\right) \sin(\phi_0 +
(\omega_J-m \omega_D)t) }{ \sqrt{(1-T_N/2) + (T_N/2)
\sum_{m=-\infty}^{\infty} J_m\left(\frac{2eV_1}{\hbar
\omega_D}\right) \cos(\phi_0 + (\omega_J-m \omega_D)t)}},
\label{acjj1}
\end{eqnarray}
\end{widetext}
where $\omega_J = 2eV_0/\hbar$ is the Josephson frequency. We note
that the $I_J$ develops a DC component whenever $ \omega_J=m_0
\omega_D$ leading to the $m_0^{\rm th}$ Shapiro step \cite{shapiro}.
The width of the step is the difference between the maximal and
minimal value of the DC component of the current. If these values
are obtained for the values $\phi_0^m$ (obtained from Eq.\
\ref{phimaxsol} by substituting $T_N \to T_N J_{m_0}[2eV_1/(\hbar
\omega_D)]$) and $-\phi_0^m$ of the relative phase, the width of the
$m_0^{{\rm th}}$ step is obtained by
\begin{eqnarray}
&& (\Delta I)_{m_0}  = \frac{I_0}{2n_1} \left(\frac{L k_F}{2
\pi}\right)^2
\int_0^{\pi/4} d\theta \int_0^{2 \pi} d \phi  \label{swidth1}\\
&& \times \frac{ (\sin 2 \theta)^{2/n_1-1} \cos 2 \theta T_N
\sin(\phi_0^m) J_{m_0}\left(\frac{2eV_1}{\hbar \omega_D}\right)}{
\sqrt{(1-T_N/2) + (T_N/2) \cos(\phi_0^m)
J_{m_0}\left(\frac{2eV_1}{\hbar \omega_D}\right)}} \nonumber
\end{eqnarray}
Using the same argument elucidated in Sec.\ \ref{and1}, we find that
for $n_1\ne n_2$ the width of the Shapiro steps are independent of
the barrier strength $\chi$ for any $m_0$. From Eq.\ \ref{swidth1},
we find that $(\Delta I)_{m_0}$ is a constant which depends on
topological winding $n_1$ and $n_2$ and on the junction geometry.
This behavior is to be contrasted with that found in conventional
JJs , where the Shapiro step width is monotonically decreasing
function of $\chi$ and with that for JJs involving 2D Dirac
materials , where they oscillate with $\chi$ \cite{ks2,yu1}. This is
shown in Fig.\ \ref{fig6}. The left panel shows that the Shapiro
step width for $m_0=1$ is independent of $\chi$ for $n_1\ne n_2$
while the right panel indicates that it has a clear oscillatory
dependence on $\chi$ for $n_1=n_2$.

Next, we consider a current biased junction in the presence of
microwave radiation since such junctions are experimentally more
relevant than their voltage-biased counterparts. Such junctions are
typically characterized by an external resistor $R$ and capacitor
$C$ connected to the JJ in parallel along with a current source. It
is well known that the equation governing the phase dynamics in such
junctions is given by \cite{likh1,yu2}
\begin{eqnarray}
\frac{d^2\phi_0}{d t^2} + \beta_0 \frac{ d \phi_0}{d t} + I_J/I_c
&=& [I_0+I_1 \cos(\omega_D t_0)]/I_c, \label{cbeq1}
\end{eqnarray}
where $I_0$ is the bias current, $I_1$ is the amplitude of the
microwave radiation, $\omega_D$ is its frequency, $I_c$ is the
critical current of the junction, $I_J$ is the Josephson current
given by Eq.\ \ref{jjcurrent} with $\phi_0 \to \phi_0(t)$, and
$\beta_0= \sqrt{\hbar/(2e I_c R^2 C)} $ is the McCumber parameter of
the junction. In Eq.\ \ref{cbeq1}, we have scaled $t \to t\omega_p$
and $\omega_D \to \omega_D/ \omega_p$, where $\omega_p^= \sqrt{2
eI_c/(\hbar C)}$ is the Plasma frequency of the JJ. The junction is
overdamped (underdamped) if $\beta_0 \gg \,(\ll) 1$. In what follows
we shall solve this equation numerically to obtain the $I-V$
characteristic of the junction following standard procedure
\cite{likh1}. We scale all current in our numerical results by $I_c$
and voltages by $V_0= \hbar \omega_p/(2e)$.

To obtain a semi-analytic understanding of the nature of $I_J$ in
current-biased Weyl junction, we provide an analytic, albeit
perturbative, solution to Eq.\ \ref{cbeq1} for $\omega_D, I_1 \gg
\beta_0$ \cite{likh1,yu2}. To this end, we note that for $I_0, I_1
\gg \beta_0, 1$, Eq.\ \ref{cbeq1} can be written as a first order
equation in $Y = d\phi_0/dt$ as
\begin{eqnarray}
\frac{d Y}{dt} + \beta_0 Y = [I_0 + I_1 \cos(\omega_D t)]/I_c
\label{obeq2}
\end{eqnarray}
The solution of this equation is straightforward and yields
\begin{eqnarray}
\phi_0(t) &=& \varphi_0 + \frac{I_0 t}{I_c \beta_0} + \frac{I_1
\sin(\omega_D t + \alpha_0)}{I_c \omega_D \gamma}, \label{phasecb1}
\end{eqnarray}
where $\gamma=\sqrt{\beta_0^2+\omega_D^2}$ and $\alpha_0 =
\arccos[\omega_D/\gamma]$. Substituting Eq.\ \ref{phasecb1} in Eq.\
\ref{acjj1} and taking note of the fact that the Shapiro steps occur
at $I_0= |m| \omega_D \beta_0 I_c$, one finds
\begin{eqnarray}
&& I_{\rm DC}[\phi_0]/I_c = \int d\theta d\phi (\sin
2 \theta)^{2/n_1 -1} \cos 2\theta \label{cbdceq1} \\
&& \times \frac{T_N J_m \left(\frac{I}{I_c \gamma \omega_D}\right)
\sin (m\alpha_0 + \varphi_0)} {\sqrt{1- \frac{T_N}{2} \left[1 - J_m
\left(\frac{I_1}{I_c \gamma \omega_D}\right) \cos(m\alpha_0 +
\varphi_0)\right]}} \nonumber
\end{eqnarray}
The width of the steps can then be obtained as in the case of the
voltage biased junction. The value of the phase $\varphi_0^{\max} +
m \alpha_0$ for which the step-size is maximal is given by Eq.\
\ref{phimaxsol} with $I_N \to T_N J_m [I_1/(I_c \gamma \omega_D)]$.
The width of the $m^{\rm th}$ Shapiro step is thus given by $2
I_{\rm DC}[\varphi_0=\varphi_0^{\rm max}]$ (Eq.\ \ref{cbdceq1}).

\begin{figure}
\rotatebox{0}{\includegraphics*[width=0.48 \linewidth]{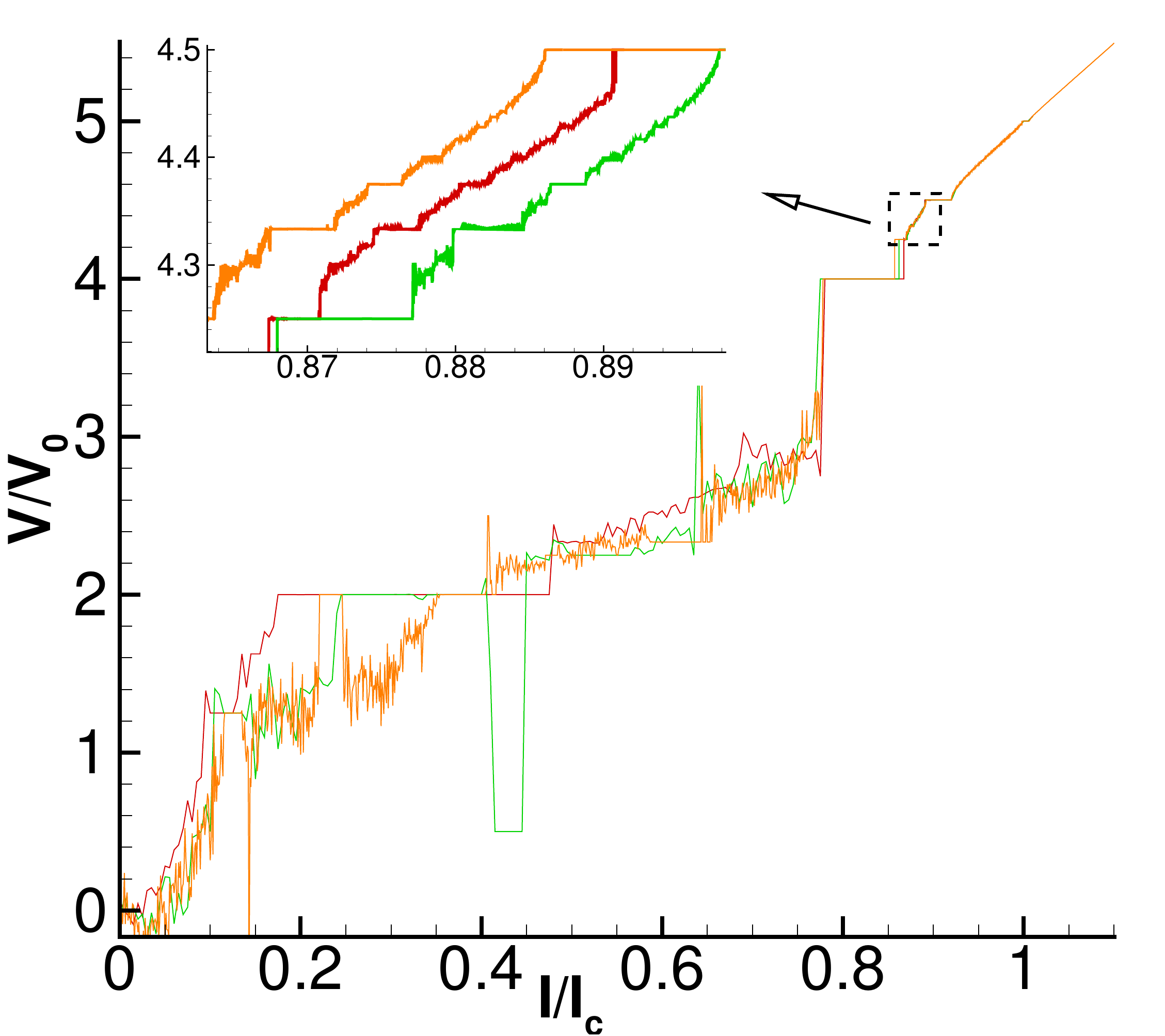}}
{\includegraphics*[width=0.48 \linewidth]{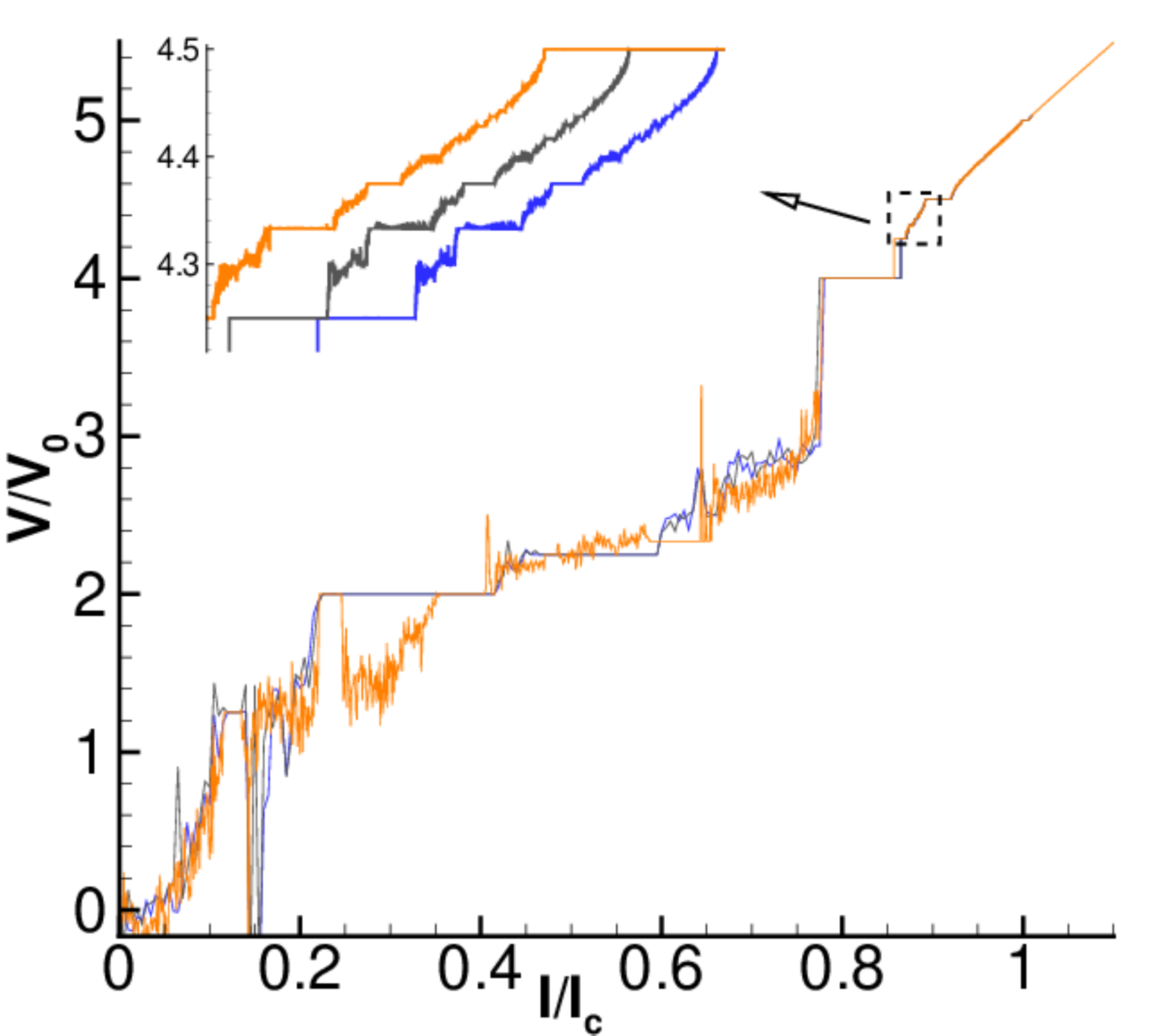}} \caption{ Plot
of the $I-V$ characteristic in the underdamped region ($\beta_0=0.2$
for $\omega_D/\omega_J= 0.5$ and $I_1/I_0=1.5$ for several
representative values of $\chi$, $n_1$ and $n_2$. The green(red)
curves in the left panel correspond to $n_1=n_2=1$ and $\chi= \pi/2
(0.01)$. The right panel corresponds to $2n_1= n_2=2$ and the
black(blue) curves correspond $\chi=\pi/2(0.01)$. The orange curves
in both correspond to $I-V$ characteristics of a conventional JJ
(where the quasiparticles obey Schrodinger equation) with
$\chi=0.01$. The inset in both panels shows clear evidence of the
devil staircase structure predicted in Ref.\ \onlinecite{yu1}. See
text for details.} \label{fig7}
\end{figure}

\begin{figure}
\rotatebox{0}{\includegraphics*[width=0.48 \linewidth]{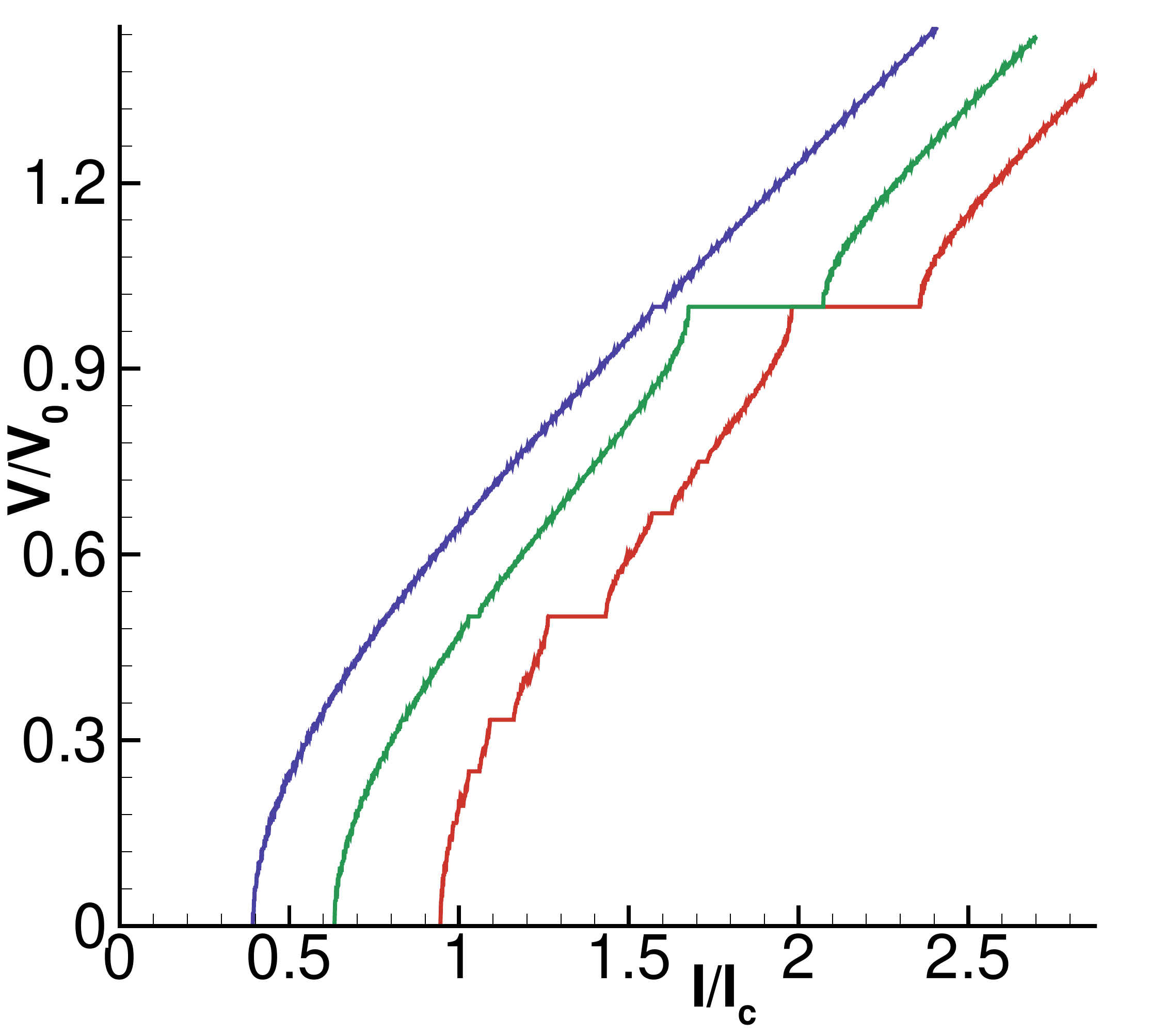}}
{\includegraphics*[width=0.48 \linewidth]{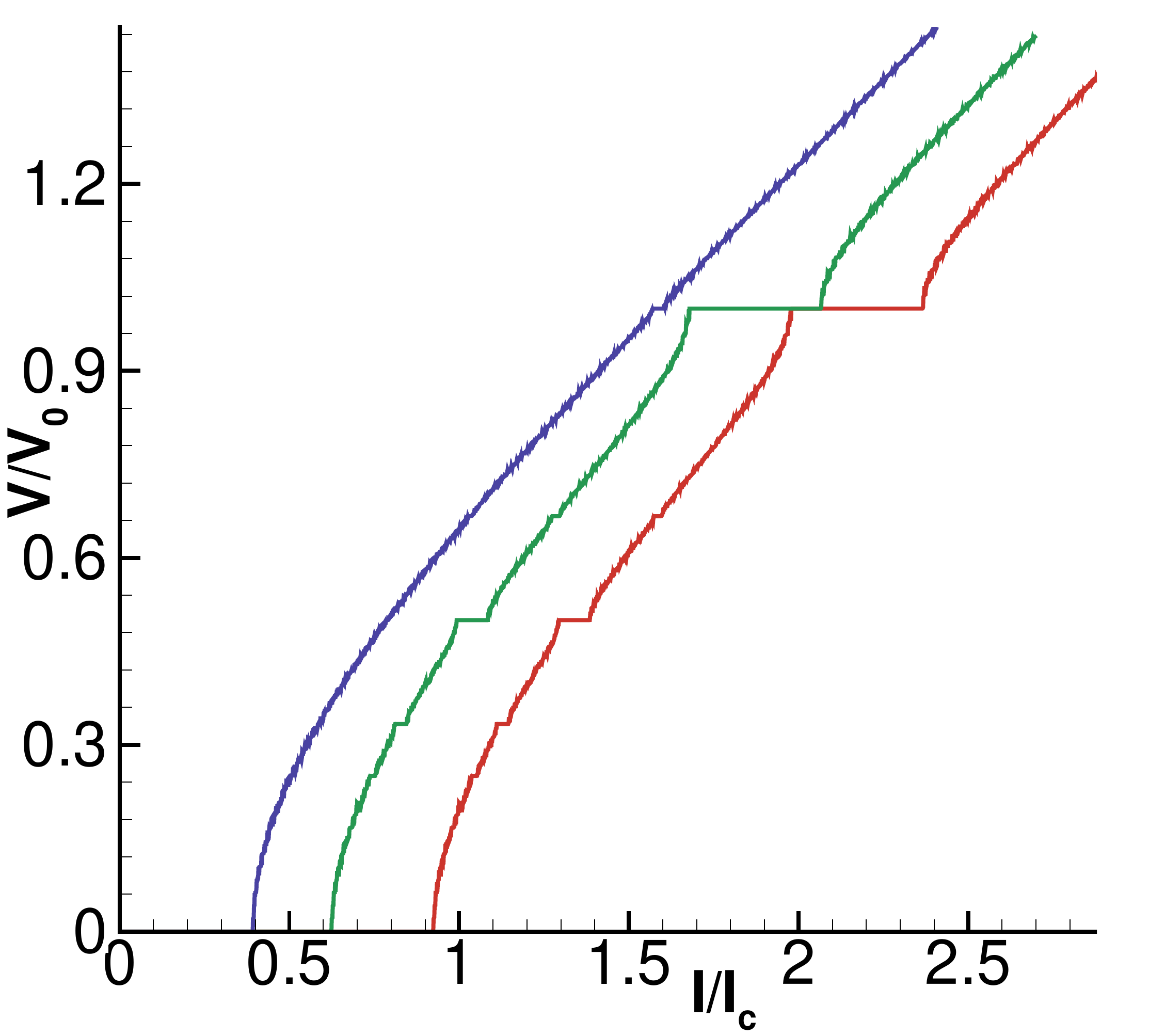}} \caption{ Plot
of the $I-V$ characteristics in the overdamped ($\beta_0=2$) region
for $n_1=n_2=1$ (left panel) and $2n_1=n_2=2$ (right panel) for
$\omega_D= \omega_J$ and $I_1=I_0$. The blue curves in both panels
correspond to conventional JJs for which the quasiparticles obey
Schrodinger equations. See text for details.} \label{fig8}
\end{figure}

\begin{figure}
\rotatebox{0}{\includegraphics*[width=0.98 \linewidth]{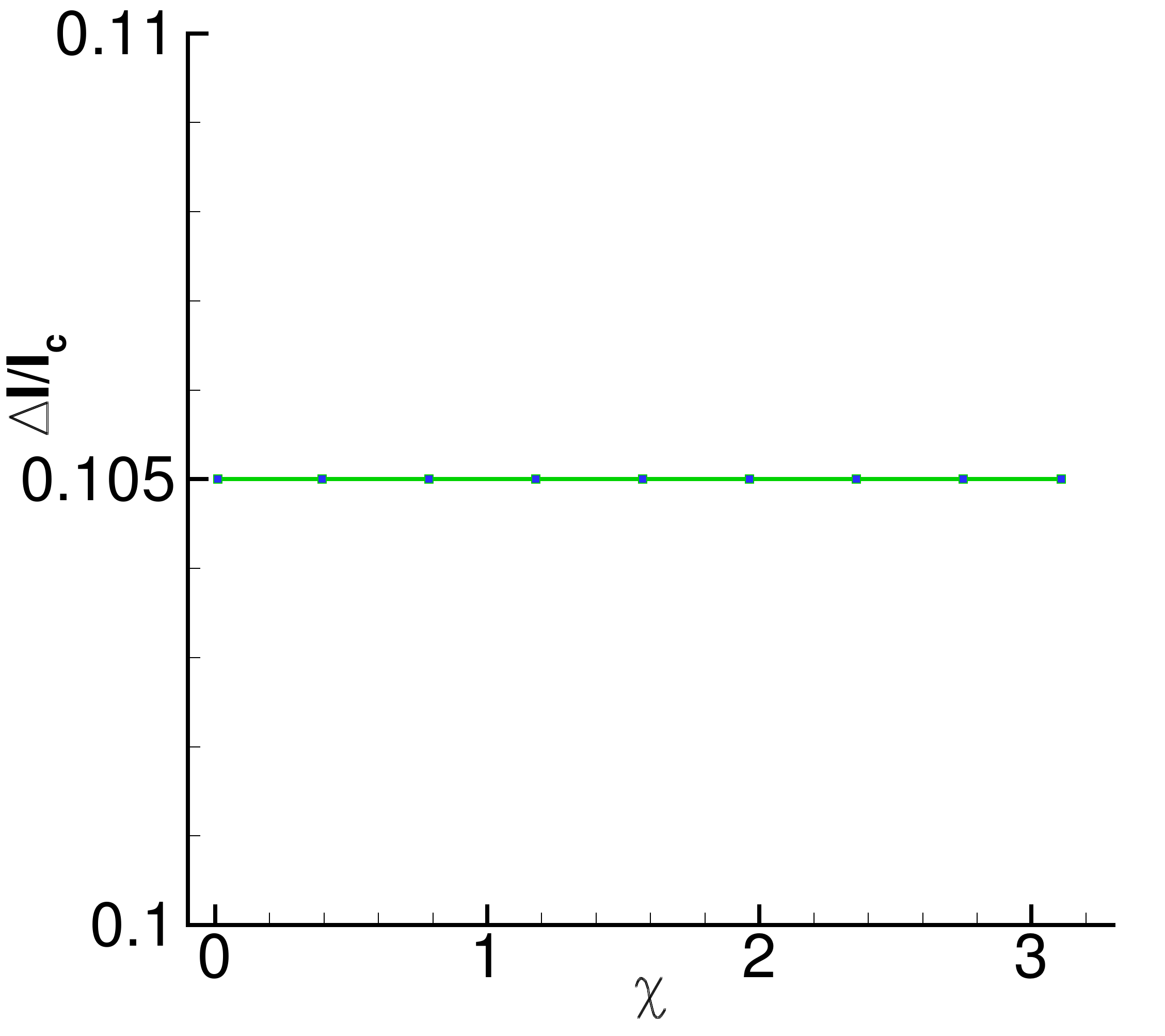}}
\caption{ Plot of the Shapiro step width computed from $I-V$
characteristics of the current biased JJ for $2n_1=n_2=2$. We have
chosen $\omega_D=3 \omega_J$ and $I_1=I_0$ and $\beta_0=0.2$. See
text for details.} \label{fig9}
\end{figure}

An exact numerical solution of Eq.\ \ref{cbeq1} leads to the $I-V$
characteristics shown in Fig.\ \ref{fig7} for the underdamped
($\beta_0=0.2$) and in Fig.\ \ref{fig8} for the overdamped
($\beta_0=2$) regimes. The left panels of both Figs.\ \ref{fig7} and
\ref{fig8} correspond to $n_1=n_2=1$ while the corresponding right
panels present data for $2n_1=n_2=2$. We find clear existence of
Shapiro steps in both the plots for small $C$. The inset in Fig.\
\ref{fig7} shows clear signature of the devil staircase structure
which is consistent with the prediction of Ref.\ \onlinecite{yu1}.
We note that the step size in the overdamped region is much larger
than that for a conventional junctions, where the quasiparticles
obey Schrodinger equation. This can be seen by the comparing the
plateaus indicated by the blue solid lines with those corresponding
to red and green lines in Figs.\ \ref{fig8}. In contrast for the
underdamped region, the step sizes are similar as can be seen from
Fig.\ \ref{fig7}. These results are in accordance with standard
expectation for such junctions \cite{likh1}. The Shapiro step widths
can be discerned from these plots for both $n_1=n_2=1$ and
$2n_1=n_2=2$ (Figs. \ref{fig7} and \ref{fig8}). A plot of these step
width as a function of $\chi$ for $\omega_D =  3 \omega_J =
6eV/\hbar$, $\beta_0=0.2$ and $I_1=I_0$ is shown in Fig.\
\ref{fig9}. The step width is found to be independent for $n_1\ne
n_2$ as per expectation. This confirms the barrier independence of
the Shapiro step width for current biased JJs involving Weyl and
multi-Weyl semimetals.

\begin{figure}
\rotatebox{0}{\includegraphics*[width=0.48 \linewidth]{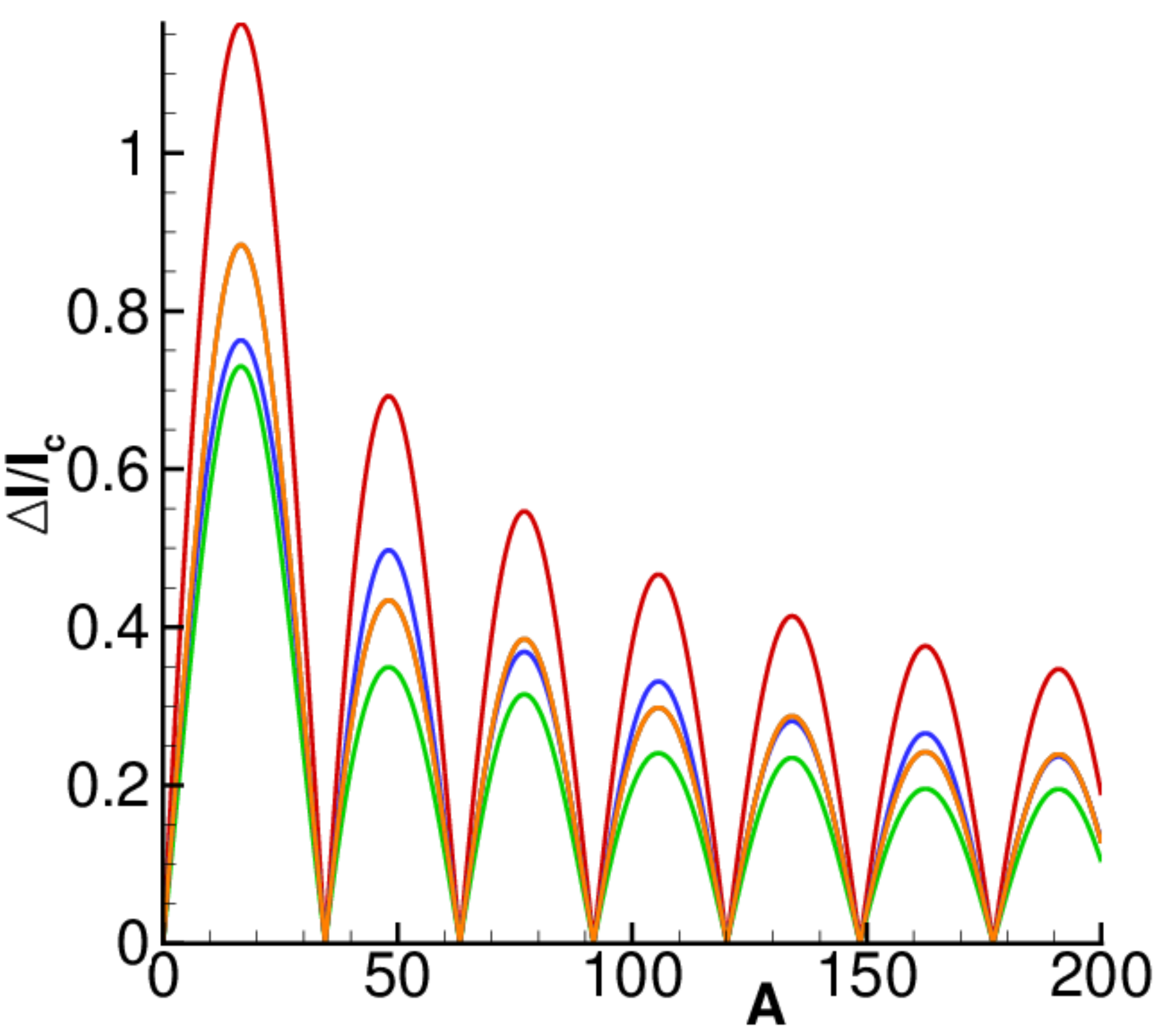}}
{\includegraphics*[width=0.48 \linewidth]{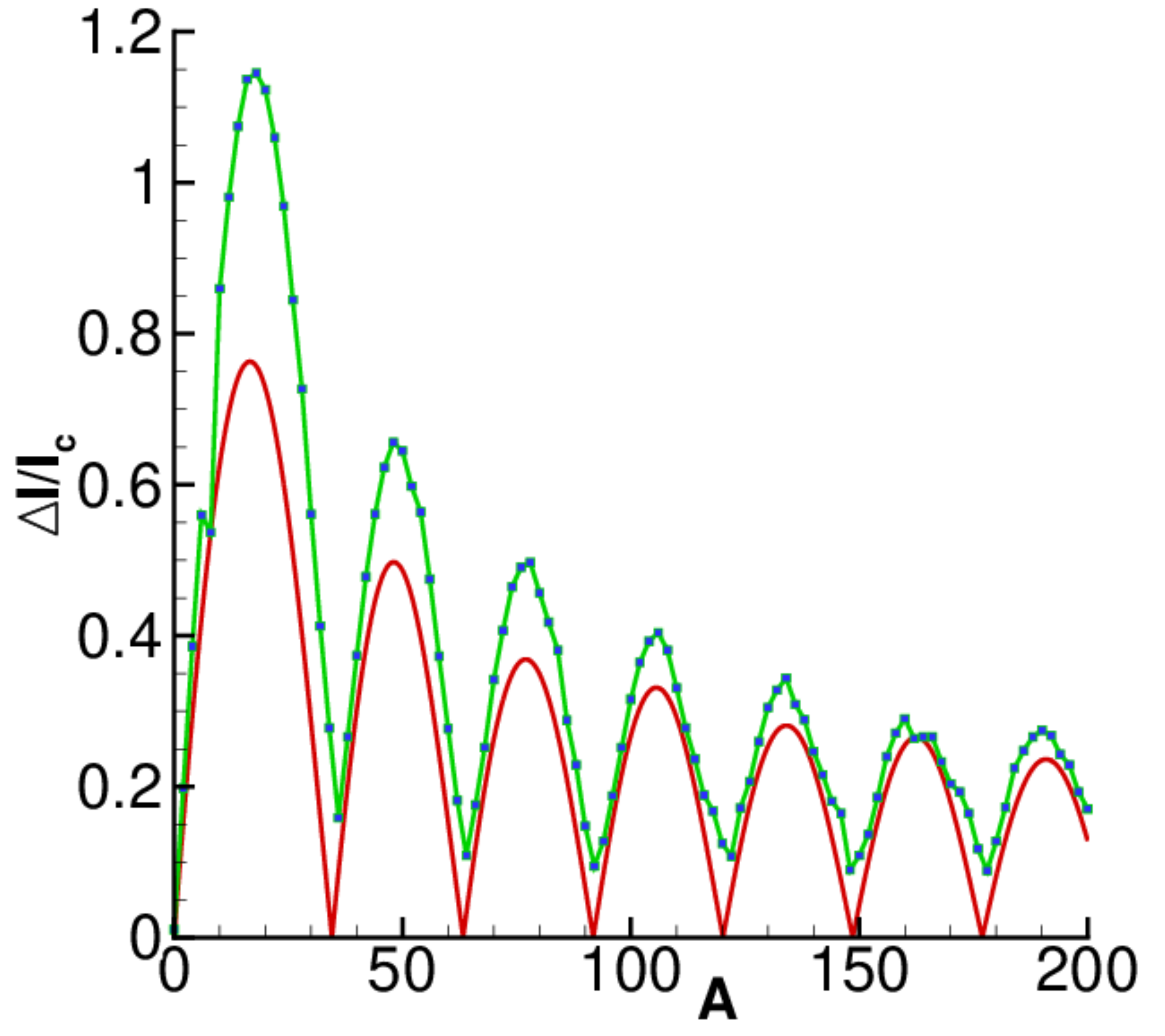}} \caption{
Left Panel: Plot of the Shapiro step width $\Delta I/I_c$ as a
function of the dimensionless radiation amplitude $A$ for several
values of $n_1$, $n_2$, and $\chi$ as obtained from Eq.\
\ref{cbdceq1}. The red curve corresponds to conventional JJ in the
AB limit, the green[blue] curve to Weyl junctions with $n_1=n_2= 1$,
and $\chi= 0.001[\pi/2]$, and the grey[orange] curve to
Weyl-multi-Weyl junction with $2n_1=n_2 = 2$, $\beta_0=0.2$, and
$\chi=0.001[\pi/2]$. The right panel shows comparison of results
obtained from semi-analytic perturbative solution (red solid line)
and exact numerics (green line with symbols) for $n_1=n_2=1$,
$\chi=0.01$, and $\beta_0=0.2$. See text for details. }
\label{fig10}
\end{figure}

Finally we study the dependence of the Shapiro step width as a
function of the dimensionless radiation amplitude $A= I_1/I_c$. This
behavior can be understood semi analytically using Eq.\
\ref{cbdceq1} for large drive amplitude. A plot of the step width,
obtained from Eq.\ \ref{cbdceq1}, is shown in the left panel of
Fig.\ \ref{fig10} for both conventional [in the AB limit] and Weyl
JJs for several representative values of $\chi$. We find that the
step-width displays an oscillatory behavior as a function of $A$;
this behavior can be understood to be the consequence of the
behavior of $J_m[A/(\gamma \omega_D)]$ as a function of $A$. The
behavior of these oscillations is therefore similar for junctions
with same or different topological winding numbers. The right panel
of Fig.\ \ref{fig10} shows the comparison of results obtained from
exact numerical solution of Eq.\ \ref{cbeq1} with those obtained
from perturbative analysis (Eqs.\ \ref{phasecb1} and
\ref{cbdceq1})for $\chi=0.01$. We find that these results agree
qualitatively even at small $A$; however, as expected, a better
quantitative agreement is achieved at large $A$.

\section{Discussion}
\label{conc}

In this work, we have studied the DC and AC Josephson effects in a
junction of Weyl-multi-Weyl semimetals. The key characteristic of
such a junction lies in the fact that it consists of junction
between two materials whose low-energy quasiparticles have {\it
different} topological winding numbers. Our results indicate that
the change in topological winding number across the junction leads
to novel features in AC and DC Josephson effects which have no
analog in conventional junctions. These features are qualitatively
distinct from their counterparts found in junction between two
topological materials with same winding number.

For DC Josephson effect, we derive an analytic expression for the
Andreev bound state in the thin barrier limit. Using it, we find
that the Josephson current $I_J$ of the junction is independent of
the dimensionless barrier strength $\chi$ in this limit. This also
allows us to show that the product $I_c R_N$ in these junctions lies
between the KO and the AB limits of conventional junctions and is
independent of $\chi$; they depend only on the topological winding
numbers $n_1$ and $n_2$ of the quasiparticles controlling the
transport in these junctions. In this sense, for a given $n_1$ and
$n_2$, $I_c R_N$ turns out to be a universal number. A deviation
from the thin barrier limit leads to oscillation of $I_c R_N$ (and
$I_J$) as a function of the barrier potential $U_0$. However, these
oscillations differ in characteristics from their counterparts in
junctions with $n_1=n_2$; their amplitude decays with increasing
$U_0$ for large $U_0/\Delta_0$ for a fixed junction width $d$. Thus
we expect the barrier independence of $I_c R_N$ to be discernible
from those in junctions of Weyl semimetals with $n_1=n_2$.

For AC Josephson effect, a study of either current or voltage biased
junctions in the presence of a microwave radiation leads to the
expected Shapiro steps. For current biased junctions, we also find
the devil staircase structure in the I-V characteristics in
accordance with earlier predictions \cite{likh1,yu1}. The width of
these Shapiro steps can be shown also to be independent of $\chi$ in
the thin barrier limit. We have also studied the variation of the
width of these steps for current biased junctions both using a
semi-analytic perturbative approach and exact numerics; these
approaches yield the expected oscillatory behavior of the step width
with amplitude of the external radiation and lead to near identical
results at large radiation amplitudes.

Our results can be verified by standard experiments used to detect
Shapiro steps. Typically such experiments are carried out with a
fixed external radiation frequency $\omega_D$; the amplitude of the
microwave radiation, or equivalently $I_1$, is varied to detect the
width of the step \cite{expshap1}. Our proposition, for junctions
between Weyl and multi-Weyl superconductors, is to carry out this
experiments with different barrier potential $\chi$. We predict that
in the thin barrier limit, the step width would be independent of
$\chi$. A deviation from the thin barrier limit would lead to
oscillations of the barrier width with $\chi$; however, the
amplitude of these oscillations would be small at high barrier
potentials which will make the behavior of these junctions distinct
from their $n_1=n_2$ counterparts.

In conclusion, we have studied JJs between Weyl and multi-Weyl
semimetals with induced $s-$wave superconductivity. We have shown
that a change of topological winding number across these junctions
ensures that their $I_c R_N$ will be independent of the bias
potential $\chi$ in the thin barrier limit. We have also shown the
independence of the Shapiro step widths on $\chi$ in both current
and voltage biased junctions in this limit. Both these properties
have no analogs in JJs of either conventional or topological
materials studied earlier and we have discussed experiments which
can detect such behavior.

\section{Acknowledgement}

The reported study was partially funded by the RFBR research
Projects No. 18-02- 00318, No. 18-32-00950 and No. 18-52-45011-IND. Numerical
calculations have been made in the framework of the RSF Project No.
18-71-10095. K.S. thanks DST, India for support through Project No.
INT/RUS/RFBR/P-314.

\end{document}